\documentclass[twocolumn,preprintnumbers,amsmath,amssymb]{revtex4}
\usepackage[dvips]{graphicx}
\usepackage{dcolumn}

\begin{document}

\title{Superfluid $^3$He in globally isotropic random media}

\date{today}

\author{Ryusuke Ikeda and Kazushi Aoyama}

\affiliation{%
Department of Physics, Graduate School of Science, Kyoto University, Kyoto 606-8502, Japan
}

\begin{abstract}
Recent theoretical and experimental studies of superfluid $^3$He in aerogels with a global anisotropy created, e.g., by an external stress, have definitely shown that the A-like phase with an equal spin pairing (ESP) in such aerogel samples is in the ABM (or, axial) pairing state. In this paper, the A-like phase of superfluid $^3$He in globally {\it isotropic} aerogel is studied in details by assuming a weakly disordered system in which singular topological defects are absent. Through calculation of the free energy, a disordered ABM state is found to be the best candidate of the pairing state of the globally isotropic A-like phase. Further, it is found through a one-loop renormalization group calculation that the coreless continuous vortices (or, vortex-skyrmions) are irrelevant to the long-distance behavior of disorder-induced textures, and that the superfluidity is maintained in spite of lack of the conventional superfluid long range order. Therefore, the globally isotropic A-like phase at weak disorder is, like in the case with a globally stretched anisotropy, a glass phase with the ABM 
pairing and showing superfluidity. 
\end{abstract}


\maketitle

\section{Introduction}

Superfluid $^3$He in aerogel, which is a random medium with a high porosity, has been studied as a prototype of impurity scattering effects on an anisotropic Cooper-paired system \cite{Bill}. The aerogel has a structural correlation length $\xi_a$, corresponding to a typical distance between neighboring strands, of the order of 30 - 50 (nm) which is comparable with the pairing coherence length $\xi_0 \simeq \hbar v_{\rm F}/(2 \pi k_{\rm B} T_c)$ in the pressure range relevant to the superfluid $^3$He. For this reason, the scattering events of the quasiparticles due to the aerogel structure are different from those of electrons in amorphous-like dirty metals corresponding to the situation with $\xi_a \ll \xi_0$ and seem to be characterized by a {\it local anisotropy} of the scattering amplitude \cite{Bill,Florida,AI05}. 

In recent years, the presence of a high pressure phase with an equal spin pairing (ESP), called the A-like phase, has become an active research subject, because it had been believed previously that the B-like phase with the Balian-Wherthamer (BW) pairing is the only stable superfluid phase in aerogel \cite{Gervais}. However, several NMR experiments have suggested the presence of the A-like phase near $T_c(P)$ and a strange lowering of the polycritical pressure (PCP) accompanying this phase \cite{Osheroff,Parpia}. Even theoretically, there have been some suggestions favoring the presence of the Anderson-Brinkman-Morel (ABM) pairing state at least at short scales \cite{TH1} by assuming a local anisotropy of aerogel structures, although the lowering of PCP has not been discussed there. On the other hand, it has been argued that the ABM pairing state in aerogel has no superfluid response at long distances, as a consequence of random orientations of ${\bf l}$-vector pinned by the local anisotropy of the aerogel and the resulting proliferation of nonsingular coreless vortices, or vortex-skyrmions \cite{Volovik96,Volovik07}, where the unit vector ${\bf l}$ is the orbital anisotropy axis and expresses the direction along which the energy gap vanishes. However, experiments seem to show nonvanishing and anisotropic superfluid responses, like in the bulk liquid \cite{Picket}. Another pairing state, called the robust pairing state, was proposed as a candidate of the A-like phase \cite{Fomin} showing superfluidity. However, it was difficult to identify this state, which is not thermodynamically stable in the bulk liquid, as the A-like phase. 

This controversy on the pairing state of the A-like phase has been resolved in more recent studies \cite{AI06,Kunimatsu,Volovik07,Erb} for the cases where the aerogel has a {\it global anisotropy} brought by an external stress. NMR measurements in both of uniaxially compressed \cite{Kunimatsu} and stretched \cite{Erb} aerogels have been nicely explained by assuming that the A-like phase in these aerogels is in the ABM pairing state with a proper alignment of the orientation of ${\bf l}$-vector. On the other hand, it seems that there is no consensus at present on the pairing state of the A-like phase in the globally isotropic case \cite{AI05,Volovik07,Fomin}. Even if the A-like phase in this case is also in the ABM pairing state, the fundamental question on the presence or absence of superfluidity in this case \cite{Volovik07} needs to be resolved. 

In the present work, thermodynamic stability of the ABM pairing state in globally isotropic random media is examined in details through comparison on the free energy between different pairing states, including the planar and robust states, by assuming some of real aerogels to be globally isotropic. Further, the presence of the quasi long range superfluid order in such a disordered ABM state is established at weak disorder where singular topological defects are absent. Here, the superfluid order will be reasonably defined through the correlation function \cite{Volovik96} 
\begin{equation}\label{corr0}
G({\bf R}) = {\rm Re} \, {\rm Tr} ( \, {\overline {{\bf \Delta}_{\bf p}({\bf r}+{\bf R}) \, {\bf \Delta}^*_{\bf p}({\bf r})}} \, ) \, , 
\end{equation}
between the spin-triplet gap parameters \cite{VW}, ${\bf \Delta}_{\bf p}({\bf r})$ and ${\bf \Delta}_{\bf p}({\bf r} + {\bf R})$. Here, the zero temperature limit is assumed so that the thermal fluctuation of the gap parameter may be neglected. Further, the overbar denotes the random average, and ${\rm Tr}$ expresses both of the trace in spin space and the average over the {\it relative} momentum ${\bf p}$ on the Fermi surface. This gap parameter, which is a tensor in spin space, depends not only on the ordinary amplitude and phase but also on the orientations of spin and orbital degrees of freedom of Cooper-pairs. 
At larger scales than the dipole coherence length \cite{VW}, the spin orientation is locked in the orbital one corresponding to the ${\bf l}$'s orientation, and a short range correlation of the ${\bf l}$'s orientation corresponds to a short range superfluid order measured by the correlation function (\ref{corr0}). The presence of a {\it quasi} long range superfluid order suggests that the corresponding superfluid correlation length is infinite. At a glance, one might wonder that such a long range correlation be destroyed by vortex-skyrmions which are generated by {\it continuous} textures of the ${\bf l}$-vector. However, we find based on a renormalization group (RG) analysis that the vortex-skyrmions appearing at short scales in globally isotropic systems may be irrelevant perturbations at long distances, implying that a nonvanishing superfluid response is well-defined. Therefore, the A-like phase at weak disorder is expected to show superfluidity, just as seen experimentally \cite{Picket}. A brief sketch on the free energy calculation in the present work has been reported elsewhere \cite{AI05} previously. 

In sec.II, the Ginzburg-Landau model including effects of randomness is derived in a form useful for a free energy calculation, and the free energy is evaluated in details in sec.III based on the Gaussian variational method often used in random systems. In sec.IV, the presence of a quasi long range superfluid order is explained by performing one-loop diagram calculations accompanying a functional RG method, and results are summarized and discussed in sec.V. Some of technical or numerical details will be explained in Appendices. 

\section{Derivation of Ginzburg-Landau action in disordered case}

As a starting microscopic Hamiltonian for deriving a Ginzburg-Landau action or functional, we choose the BCS Hamiltonian with an attractive interaction in the purely $p$-wave channel, which is written in the familiar notation as 
\begin{equation}
{\hat H}_{p} - \mu {\hat N} = \sum_{{\bf p}, \alpha} \, \xi_{\bf p} \, {\hat a}^\dagger_{{\bf p},\alpha} \, {\hat a}_{{\bf p},\alpha} + {\hat H}_{\rm int}, 
\end{equation}
where 
\begin{eqnarray}\label{hamil0}
{\hat H}_{\rm int} &=& - 3|g| \sum_{\bf q} {\hat O}_{\mu,j}^\dagger({\bf q}) \, {\hat O}_{\mu,j}({\bf q}), \nonumber \\
{\hat O}_{\mu,j}({\bf q}) &=& \sum_{\bf p} \frac{p_j}{2 p_F} \, {\hat a}_{-{\bf p}+{\bf q}/2, \alpha} ({\rm i} \, \sigma_\mu \, \sigma_2)_{\alpha \beta} \, {\hat a}_{{\bf p}+{\bf q}/2, \beta}.
\end{eqnarray}
Performing the standard decoupling \cite{Popov} in ${\hat H}_{\rm int}$ by introducing the pair-field $A_{\mu,j}$, where $\mu$ ($j$) denotes the 3-components of the spin (orbital) degree of freedom, the superfluid part of the partition function is given by $\langle T_s \exp[\, - \int^{1/T}_0 ds {\hat H}_{\rm int}(s) \, ] \rangle = \int {\cal D}\Delta {\cal D}\Delta^* \exp(-{\cal S})$ in the $\hbar=k_{\rm B}=1$ unit, where 
\begin{eqnarray}\label{SGL}
{\cal S} = \sum_{\bf q} \frac{1}{3 \, |g| T} \, A^*_{\mu,j}({\bf q}) A_{\mu,j}({\bf q})  - 
{\rm ln} \langle T_s \exp{\Pi} \rangle, 
\end{eqnarray}
\begin{eqnarray}
\Pi &=& \! \frac{1}{2} \sum_{\bf q} \int_{\bf p} 
\biggl[ ( \Delta^\dagger_{\hat p} ({\bf q}) )_{\beta, \alpha} 
\int_0^{T^{-1}} ds \, {\hat a}_{p+q/2, \alpha}(s) \, 
{\hat a}_{-p+q/2, \beta}(s) 
\biggr] \nonumber \\
&+& {\rm h.c.}, 
\end{eqnarray} 
$\int_{\bf p}$ denotes the momentum integral $\int d^3p/(2 \pi)^3$, $(\Delta_{\hat p}({\bf q}))_{\alpha, \beta} = A_{\mu,i}({\bf q}) \, {\hat p}_i ({\rm i} \sigma_\mu \sigma_2)_{\alpha, \beta}$ is the pair-field, $\psi_\sigma({\bf r}) = \sum_p {\hat a}_{p, \sigma} e^{{\rm i}{\bf p}\cdot{\bf r}}$ is the quasiparticle field, and $\langle \, \, \rangle$ expresses the ensemble average over the quasiparticle distribution. The GL action is obtained by, in ${\cal S}$, keeping just the quadratic and quartic terms in $A_{\mu,j}$. The pair-field is assumed to be independent of the imaginary time $s$ because quantum fluctuations of $A_{\mu,i}$ do not have to be included in considering superfluid phases of $^3$He in equilibrium, in which fluctuation effects are safely negligible. For $^3$He in aerogel, the total quasiparticle Hamiltonian needs to include a term associated with an impurity scattering. As usual, it will be expressed hereafter as a nonmagnetic 
random potential term \cite{comhe4}
${\cal H}_{\rm imp} = \sum_\sigma \int d^3r \, u({\bf r}) \psi^\dagger_\sigma({\bf r}) \psi_\sigma({\bf r})$. 
The scattering potential $u({\bf r})$ has zero mean, and the quasiparticle life time $\tau$ is defined by the relation $\tau^{-1}= 2 \pi N(0) \langle {\overline {|u_{{\bf p}-{\bf p}'}|^2}} \rangle_{\hat p}$, where $N(0)$ is the density of states per spin in the normal state, and 
$\langle \,\,\, \rangle_{\hat p}$ denotes the angle-average over the orientation of the relative momentum ${\bf p}$ on the Fermi surface. If the aerogel we assume has no global anisotropy, $\tau$ defined above is independent of the quasiparticle momentum ${\bf p}'$. Using a quasiparticle Green's function \cite{AGD} $G_{\varepsilon} ({\bf p}, {\bf p}')$ defined {\it prior to} the impurity average, the quadratic part ${\cal S}_2$ of ${\cal S}$ is expressed as 
\begin{eqnarray}\label{S2}
{\cal S}_2 &=& T^{-1} \sum_{q,q'} \biggl[ \frac{\delta_{i,j} \delta_{q,q'}}{3 |g|} - T \sum_{\varepsilon} \int_{\bf p} \int_{{\bf p}'} {\hat p}_i {\hat p}_j \nonumber \\
&\times& {\overline {G_{\varepsilon} ({\bf p}+{\bf q}/2, {\bf p}'+{\bf q}'/2) \, G_{-\varepsilon} (-{\bf p}+{\bf q}/2, -{\bf p}'+{\bf q}'/2)}} \biggr] \nonumber \\
&\times& A_{\mu,i}^*({\bf q}) A_{\mu,j}({\bf q}'). 
\end{eqnarray}
In the present situation where the critical fluctuation is negligible, a ${\bf q}$-dependence of $A_{\mu,j}$ follows from the quenched disorder. In the GL regime where the amplitude of $A_{\mu,j}$ is small, it is sufficient to keep, in ${\cal S}_2$, disorder-induced terms related to the ${\bf q}$-dependences of $A_{\mu,j}$, and the corresponding contributions from the GL quartic term may be neglected. Then, the quartic term in our GL action takes the same form as the familiar one for the disorder-free bulk liquid $^3$He (see, e.g., Ref.\cite{VW}) 
\begin{eqnarray}\label{S4}
{\cal S}_4 &=& T^{-1} \sum_{q_1,q_2,q_3} (\beta_1 |A_{\mu,i} A_{\mu,i}|^2 + \beta_2 (A_{\mu,i}^* A_{\mu,i})^2 \nonumber \\
&+& \beta_3 A^*_{\mu,i} A^*_{\nu,i} A_{\mu,j} A_{\nu,j} + \beta_4 A^*_{\mu,i} A_{\nu,i} A^*_{\nu,j} A_{\mu,j} \nonumber \\
&+& \beta_5 A^*_{\mu,i} A_{\nu,i} A^*_{\mu,j} A_{\nu,j} ).
\end{eqnarray}
In the weak coupling limit without any vertex correction, ${\cal S}_4$ is obtained from the expression 
\begin{eqnarray}\label{S4wc} 
{\cal S}_{4, {\rm wc}} &\simeq& \sum_{q_j, {\varepsilon}} \int_{\bf p} (\, G_{\varepsilon}({\bf p}) \, G_{-\varepsilon}(-{\bf p}) \,)^2 
 {\rm Tr}(\Delta_{\hat p}^\dagger \Delta_{\hat p} \Delta^\dagger_{\hat p} \Delta_{\hat p})  
\end{eqnarray}
expressing the Gor'kov box of Fig.1(a), 
where $G_{\varepsilon}({\bf p}) = ({\rm i} {\tilde \varepsilon} - \xi_{\bf p})^{-1}$ with ${\tilde \varepsilon}=\varepsilon + {\rm sgn}(\varepsilon)/(2 \tau)$ is the impurity-averaged quasiparticle propagator. The resulting ${\cal S}_{4, {\rm wc}}$ is given by replacing $\beta_j$ in eq.(\ref{S4}) by $\beta_j^{({\rm wc})}$, where $\beta^{({\rm wc})}_2 = \beta^{({\rm wc})}_3 = \beta^{({\rm wc})}_4 = - \beta^{({\rm wc})}_5 = -2 \beta^{({\rm wc})}_1= 2 \beta^{({\rm wc})}(T)$, 
\begin{eqnarray}\label{betawc}
\beta^{({\rm wc})}(T) &=& \frac{N(0)}{240 \pi^2 T^2} \sum_{n=0} \frac{8}{(2n+1+(2 \pi T \tau)^{-1})^3} \nonumber \\ 
&\equiv& \frac{\beta_0(T)}{7 \zeta(3)} \sum_{n=0} 
\frac{8}{(2n+1+(2 \pi T \tau)^{-1})^3}, 
\end{eqnarray}
and $\zeta(3) \simeq 1.2$. 
\begin{figure}[t]
\scalebox{1.5}[1.5]{\includegraphics{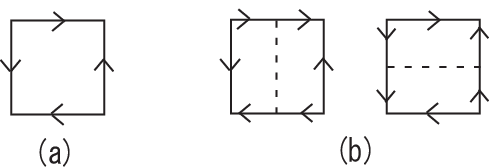}}
\caption{Diagrams expressing Gor'kov boxes leading to eqs.(9) and (10). The solid line with arrow and the dashed line denote the quasiparticle Green's function and the impurity line carrying $\tau^{-1}$, respectively.}
\label{fig:Gorkov}
\end{figure}

In performing the impurity average in eq.(\ref{S4wc}), the leading term in $(E_{\rm F} \tau)^{-1}$ was kept by neglecting diagrams with crossed impurity lines \cite{AGD}. We need to comment on our neglect in eq.(\ref{S4wc}) of two types of vertex corrections induced by the impurities. First, the impurity-ladders dressing the four 
corners of Fig.1(a) were neglected. These vertex corrections of the order of $1/(2 \pi T \tau)$ are present even in the $p$-wave pairing case because we take account of wave vector-dependences of the squared impurity potentials $|u_{\bf k}|^2$. However, they merely renormalize the magnitude of $\beta_0$ and never affects a {\it relative difference} between free energies of two different 
pairing states. On the other hand, irrespective of the pairing interaction, there are also additional diagrams, described in Fig.1(b), accompanied by a single impurity line. These diagrams do not contribute to $\beta^{({\rm wc})}_1$ and $\beta^{({\rm wc})}_3$, while they change other coefficients as follows \cite{TH1,AI07} : $\beta^{({\rm wc})}_2 \to \beta^{({\rm wc})}_2 - \beta_0 \Delta_{\rm imp}$, $\beta^{({\rm wc})}_4 \to \beta^{({\rm wc})}_4 - \beta_0 \Delta_{\rm imp}$, and $\beta^{({\rm wc})}_5 \to \beta^{({\rm wc})}_5 + \beta_0 \Delta_{\rm imp}$, where $\Delta_{\rm imp}$ is of the order $(2 \pi T \tau)^{-1} \ll 1$. 


As is well known, the coefficients $\beta_j$ ($j=1$ to $5$) appear in a manner dependent on the pairing state in the condensation energy of bulk $^3$He. Hereafter, such $\beta_j$'s combinations in the ABM, BW, planar, and the robust states are denoted by $\beta_{\rm ABM}$, $\beta_{\rm BW}$, $\beta_{\rm P}$, and $\beta_{\rm R}$, respectively \cite{com2} which will be indicated as $\beta_{\rm N}$ in the lump hereafter. In the disordered case, they are expressed as 
\begin{eqnarray}\label{betazoo}
\beta_{\rm ABM} &=& \beta_{245} - \beta_0 \Delta_{\rm imp}, \nonumber \\
\beta_{\rm BW} &=& \beta_1 + \beta_2 + \frac{\beta_{345}}{3} 
- \beta_0 \Delta_{\rm imp}, \nonumber \\
\beta_{\rm P} &=& \beta_1 + \beta_2 + \frac{\beta_{345}}{2} 
- \beta_0 \Delta_{\rm imp}, \nonumber \\
\beta_{\rm R} &=& \beta_2 + \frac{1}{9}(\beta_{13} + 5 \beta_{45})-\beta_0 \Delta_{\rm imp},
\end{eqnarray}
where $\beta_{ij}=\beta_i+\beta_j$, and $\beta_{ijk}=\beta_i + \beta_j + \beta_k$. 
Note that $\Delta_{\rm imp}$ appears in the {\it same} form in all $\beta_{\rm N}$s. Further, a pairing state with a small $\beta_{\rm N}$ has a lower free energy at the mean field level. Therefore, relative stability between the different pairing states cannot be reversed by including the contribution of Fig.1(b). 

Besides, $\beta_j$ must include the so-called strong coupling (SC) corrections [21,22,19] 
which, in clean limit, stablize the ABM state as the bulk A-phase of superfluid $^3$He. In the disordered case, the SC corrections to $\beta_j$ consist of two contributions. One is the expression in clean limit with the Matsubara frequency $\varepsilon$ replaced by ${\tilde \varepsilon}$. The other consists of terms including impurity-induced vertex corrections. Hereafter, they will be denoted by $\delta \beta_j^{({\rm sc})}$ and $\delta {\tilde \beta}_j^{({\rm sc})}$, respectively \cite{AI07}. That is, we have 
$\beta_j=\beta_j^{({\rm wc})} + \delta \beta_j^{({\rm sc})} + \delta {\tilde \beta}_j^{({\rm sc})}$. 
Details of $\delta \beta_j^{({\rm sc})}$ and $\delta {\tilde \beta}_j^{({\rm sc})}$ were examined in Ref.\cite{AI07} thoroughly. Their pressure dependences in each pairing state are needed in obtaining a theoretical phase diagram, and their numerical values will be illustrated for reference in Appendix A. 

Now, let us turn to detailing the second term of ${\cal S}_2$ 
by expanding $f_{ij}(\varepsilon) \equiv \int_{\bf p} \int_{{\bf p}'} {\hat p}_i {\hat p}'_j G_{\varepsilon} ({\bf p}_+, {\bf p}'_+) \, G_{-\varepsilon} (-{\bf p}_-, -{\bf p}'_-)$ in powers of the impurity potential $u$, where ${\bf p}_\pm = {\bf p} \pm {\bf q}/2$. Up to O($u^2$), we express 
$f_{ij}$ as  $f^{(0)}_{ij} + f^{(1)}_{ij} + f^{(2)}_{ij}$, where 
\begin{eqnarray}\label{S2dis0}
f^{(0)}_{ij}(\varepsilon) = \frac{\delta_{i,j}}{3} \delta_{{\bf q}, {\bf q}'} \int_{\bf p} G_{\varepsilon}({\bf p}_+) G_{-\varepsilon}(-{\bf p}_-), 
\end{eqnarray}
\begin{eqnarray}\label{S2dis1}
f^{(1)}_{ij}(\varepsilon) &=& - \int_{\bf p} {\hat p}_i {\hat p}_j  G_{\varepsilon}({\bf p}_+) ( G_{-\varepsilon}(-{\bf p}_-) G_{-\varepsilon}(-{\bf p}_+ + {\bf q}') \nonumber \\
&+& G_{-\varepsilon}(-{\bf p}_-) G_{\varepsilon}({\bf p}_- + {\bf q}') ) u_{{\bf q}-{\bf q}'}, 
\end{eqnarray}
and 
\begin{widetext}
\begin{eqnarray}\label{S2dis2}
f^{(2)}_{ij}(\varepsilon) &=& k_{\rm F}^{-2} \int_{\bf p} \int_{\bf k} \biggl[ \biggl({\bf p} + \frac{\bf k}{2} \biggr)_i \biggl({\bf p} - \frac{\bf k}{2} \biggr)_j G_\varepsilon({\bf p}+{\bf k}/2) G_\varepsilon({\bf p}-{\bf k}/2) G_{-\varepsilon}(-{\bf p}-{\bf k}/2) G_{-\varepsilon}(-{\bf p}+{\bf k}/2) u_{{\bf k}+{\bf q}} u_{-{\bf k}-{\bf q}'} \nonumber \\ 
&+& \biggl({\bf p} + \frac{\bf k}{2} \biggr)_i \biggl({\bf p} + \frac{\bf k}{2} \biggr)_j \biggl( G_\varepsilon({\bf p}+{\bf k}/2) (G_{-\varepsilon}(-{\bf p}-{\bf k}/2))^2 G_{-\varepsilon}(-{\bf p}+{\bf k}/2) \nonumber \\
&+& ( G_\varepsilon({\bf p}+{\bf k}/2))^2 G_{\varepsilon}({\bf p}-{\bf k}/2) G_{-\varepsilon}(-{\bf p}-{\bf k}/2) \biggr) 
u_{{\bf k}+{\bf q}} u_{-{\bf k}-{\bf q}'} \biggr]. 
\end{eqnarray}
\end{widetext}
The contributions in ${\cal S}_2$ corresponding to eqs.(\ref{S2}) and (\ref{S2dis0}) give the ordinary quadratic term in the so-called Abrikosov-Gor'kov approximation and in the weak-coupling limit \cite{AG,TH1}. Its expression is well known and given by 
\begin{eqnarray}
{\cal S}_2^{(0)} &=& T^{-1} \sum_{q} \biggl[ \alpha \delta_{i,j} + \frac{1}{2} \biggl( 2 K_1 q_i q_j + K_2 q^2 \delta_{i,j} \biggr) \biggl] \nonumber \\
&\times& A^*_{\mu,i}({\bf q}) A_{\mu,j}({\bf q}),   
\end{eqnarray}
where 
\begin{eqnarray}
\alpha &=& \frac{N(0)}{3} \biggl[ {\rm ln}\biggl(\frac{T}{T_{c0}} \biggr) + \psi(1/2 + 1/[4 \pi T \tau]) - \psi(1/2) \biggr], \nonumber \\
K_1 &=& K_2 = \frac{2}{5} N(0) \xi_0^2, 
\end{eqnarray}
$T_{c0}$ is the superfluid transition temperature of the bulk liquid, $\psi(z)$ is the digamma function, and 
\begin{equation}
\xi_0 = \frac{v_{\rm F}}{2 \pi T} \sqrt{\frac{1}{12} \sum_{n \geq 0} (n+1/2+ 1/(4 \pi T \tau))^{-3}}
\end{equation}
is the coherence length. 

In $f_{ij}$, the first order term $\sum_{\varepsilon} f^{(1)}_{ij}(\varepsilon)$ is easily found to vanish after performing the ${\bf p}$-integral. 
Thus, we have only to focus on $f^{(2)}_{ij}$. After substituting $f^{(2)}_{ij}$ into eq.(\ref{S2}), a larger $|{\bf k}|$ is found to become dominant in the resulting replicated action ${\overline {\cal S}}_{\rm dis}$ (see below), while, for $|{\bf q}|$, $|{\bf q}'| < 2 \pi \xi_0^{-1}$, any ${\bf q}$ and ${\bf q}'$ dependences included in the ${\bf p}$-integral are negligible compared to the large $|{\bf k}|$. Then, the ${\bf p}$-integral in $f^{(2)}_{ij}$ is derived in the conventional manner \cite{AGD} used for obtaining the static supefluid response, and we obtain 
\begin{equation}
f^{(2)}_{ij}(\varepsilon) \simeq - \frac{\pi^2}{8} \int_{\bf k} {\hat k}_i {\hat k}_j \frac{N(0)}{E_{\rm F}{\tilde \varepsilon}^2} \, \frac{k_F}{|{\bf k}|} \, u_{{\bf k}+{\bf q}} u_{-{\bf k}-{\bf q}'}. 
\end{equation}
\begin{figure}[t]
\scalebox{1.5}[1.5]{\includegraphics{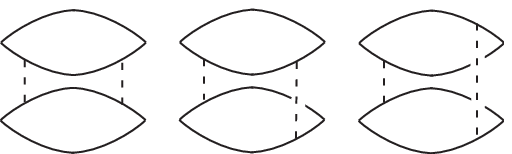}}
\caption{Diagrams giving ${\overline {\cal S}}_{\rm dis}$.}
\label{fig:disdiag}
\end{figure}
Although, by substituting this into ${\cal S}_2$, a disorder-induced term appears in the GL action, it is more convenient to directly work in the corresponding quartic term ${\overline {\cal S}}_{\rm dis}$ arising {\it after} the impurity-average of the free energy, where the index "dis" implies "disorder". To do this, let us introduce the replica-description of the averaged free energy \cite{EA} 
\begin{eqnarray}
{\overline F} = -T \lim_{n \to +0} \frac{{\overline {Z^n}} - 1}{n},
\end{eqnarray} 
where ${\overline {Z^n}} = {\rm Tr} \exp(-{\overline {\cal S}})$. The quartic term ${\overline {\cal S}}_{\rm dis}$, described in Fig.2, appears in the replicated GL action ${\overline {\cal S}}$ and is given by \cite{com3}
\begin{widetext}
\begin{eqnarray}
{\overline {\cal S}}_{\rm dis} &=& - T^{-2} \sum_{a,b=1}^{n}  \int_{\bf k} {\hat k}_i {\hat k}_j {\hat k}_r {\hat k}_s \biggl(\frac{N(0) \, k_F}{16 E_{\rm F} T |{\bf k}|} \psi^{(1)}\biggl(\frac{1}{2}+\frac{1}{4 \pi T \tau} \biggr) \biggr)^2 \sum_{a,b} \sum_{q_1,q_2, q_3} {\overline {|u_{{\bf k}+{\bf q}_1} u_{{\bf k}+{\bf q}_2}|^2}} (A^{(a)}_{\mu,i}(q_1))^* (A^{(b)}_{\nu,r}(q_3))^* \nonumber \\
&\times& A^{(a)}_{\mu,j}(q_2) A^{(b)}_{\nu,s}(q_1+q_3-q_2) \nonumber \\
 &\simeq& - T^{-1} \frac{\pi^2}{960} \frac{T}{E_{\rm F}} \frac{N(0)}{T^2} \frac{\gamma}{(\tau T)^2} \biggl(\psi^{(1)}\biggl(\frac{1}{2}+\frac{1}{4 \pi T \tau} \biggr) \biggr)^2 \sum_{a,b} \sum_{q_1,q_2, q_3} (\delta_{i,j} \delta_{r,s} + \delta_{i,r} \delta_{j,s} + \delta_{i,s} \delta_{r,j} ) (A^{(a)}_{\mu,i}(q_1))^* (A^{(b)}_{\nu,r}(q_3))^* \nonumber \\
&\times& A^{(a)}_{\mu,j}(q_2) A^{(b)}_{\nu,s}(q_1+q_3-q_2),
\end{eqnarray}
\end{widetext}
where $\psi^{(1)}(z)=d\psi(z)/dz$, and 
\begin{equation}\label{locani}
\gamma \equiv (\tau N(0))^2 \int_0 \frac{dk}{2 \pi^2 k_{\rm F}} {\overline {|u_{\bf k}|^4}}.
\end{equation}
It is easy to verify that ${\overline {\cal S}}_{\rm dis}$ can also be obtained by assuming the following quadratic action to be present in the original action ${\cal S}$ :
\begin{eqnarray}
{\cal S}_{2,{\rm dis}} &=& \int d^3r [ \, U({\bf r}) \delta_{i,j} + V({\bf r}) (\delta_{i,j} -3 {\hat a}_i({\bf r}) {\hat a}_j({\bf r}) ) \, ] \nonumber \\
&\times& A^*_{\mu, i}({\bf r}) A_{\mu,j}({\bf r})
\end{eqnarray}
Here, ${\hat a}_i$ yields a Gaussian ensemble satisfying ${\overline {{\hat a}_i}}=0$ and $3{\overline {{\hat a}_i({\bf r}) {\hat a}_j({\bf r})}} = \delta_{i,j}$, while the potentials $U$ and $V$ have zero mean and satisfy ${\overline {U({\bf r}) V({\bf r}')}}=0$, and 
\begin{eqnarray}
{\overline {U({\bf r}) U({\bf r}')}} &=& {\overline {V({\bf r}) V({\bf r}')}} 
= T^{-1} \delta \beta_{\rm d} \delta^{(3)}({\bf r}-{\bf r}')
\end{eqnarray}
with 
\begin{eqnarray}
\delta\beta_{\rm d} = \beta_0 \frac{\gamma}{E_{\rm F} T \tau^2} \frac{5 \pi^4}{42 \zeta(3)} \biggl( \psi^{(1)}\bigl(\frac{1}{2}+\frac{1}{4 \pi T \tau} \bigr) \biggr)^2. 
\end{eqnarray}
In this way, one can regard the {\it original} GL action ${\cal S}$ 
below $T_c$ 
as
\begin{equation}
{\cal S} = {\cal S}_2^{(0)} + {\cal S}_4 +  {\cal S}_{2, {\rm dis}}. 
\end{equation}

Hereafter, the pair-field $A_{\mu,i}$ will be expressed by separating the amplitude $|\Delta|$ from the symmetry variables \cite{VW} composed of the spin and orbital degrees of freedom together with the overall phase $\Phi$ in the manner 
\begin{equation}
A_{\mu,i} = |\Delta| a_{\mu,i}. 
\end{equation}
Following the standard notation, $a_{\mu,i}$ in the ABM state is given by 
\begin{equation}\label{abmtext}
a_{\mu,i} = e^{{\rm i} \Phi} \, \frac{d_\mu \, ({\bf m} + {\rm i} \, {\bf n})_i}{\sqrt{2}}
\end{equation} 
with the triad (${\bf m}$, ${\bf n}$, ${\bf l}$) of unit vectors. On the other hand, it takes the form, $e^{{\rm i}\Phi} \, R_{\mu,i}/\sqrt{3}$ and $e^{{\rm i}\Phi} \, R_{\mu,k} \, \delta^{\rm T}_{i,k}/\sqrt{2}$, for the BW and planar states, respectively, where $R_{\mu,i}$ is a rotation matrix, 
and $\delta^{\rm T}_{i,j}=\delta_{i,j} - {\bf l}_i {\bf l}_j$ \cite{VW}. According to the definition of the ${\bf l}$-vector mentioned in sec.I, the same notation on the anisotropy axis will be used for both of the ABM and planar 
states. 
Then, ${\overline {\cal S}}_{\rm dis}$ is rewritten for the ABM pairing state in the form ${\overline {\cal S}}_{{\rm dis}(1)}+{\overline {\cal S}}_{{\rm dis}(2)}$, where 
\begin{equation}
{\overline {\cal S}}_{{\rm dis}(1)} = - \frac{1}{2T} \frac{6}{5} \delta\beta_{\rm d} \sum_{a,b} \int d^3r |\Delta^{(a)}({\bf r}) \Delta^{(b)}({\bf r})|^2, 
\end{equation} 
and 
\begin{eqnarray}\label{sdis2}
{\overline {\cal S}}_{{\rm dis}(2)} &=& - \frac{1}{2T} \frac{3}{10} \delta\beta_{\rm d} \sum_{a,b} \int d^3r |\Delta^{(a)}({\bf r}) \Delta^{(b)}({\bf r})|^2 \nonumber \\
&\times& [({\bf l}^{(a)}\cdot{\bf l}^{(b)})^2 - 1]. 
\end{eqnarray} 
The corresponding action for the planar state takes the same form as above, while ${\overline {\cal S}}_{\rm dis}$ is given, in the BW and robust states, simply by $5 \, {\overline {\cal S}}_{{\rm dis}(1)}/6$. Since eq.(\ref{sdis2}) is nonvanishing only if the ${\bf l}$-vector is spatially varying so that $({\bf l}^{(a)} \cdot {\bf l}^{(b)})^2 - 1$ is nonzero, $|\Delta^{(a)}({\bf r}) \Delta^{(b)}({\bf r})|^2$ in eq.(\ref{sdis2}) may be replaced by its mean field value $|\Delta_{\rm MF}|^4$ below the critical region in the close vicinity of $T_c$, or as far as a slowly varying $a_{\mu,i}$ is assumed.  Of course, $|\Delta_{\rm MF}|^2$ needs to be determined by examining ${\overline F}$. 

Before ending this section, it will be appropriate to discuss about the treatment on the impurity scattering used in this paper. Our procedure on the impurity scatterings used in ${\cal S}_2^{(0)}$ and ${\cal S}_4$ is more or less an extension of the Abrikosov-Gor'kov approach \cite{AG} based on the Born approximation. Except in the situation with extremely weak disorder, the so-called unitary limit including multiple scattering processes is often used by assuming the {\it isotropic} $s$-wave scattering event to be dominant (see, e.g., Ref.\cite{TH1}). On the other hand, an inhomogenuity of the order parameter amplitude $|\Delta|$ to be created spontaneously \cite{Franz} by impurity scatterings was not incorporated in the present analysis. In an {\it isotropic} approximation, this effect becoming more important at higher pressures was studied in an elaborate numerical work \cite{Th2}. Throughout the present paper, however, we argue that the local or global anisotropy of scattering events in aerogel, which has not been incorporated in calculations in previous microscopic works \cite{TH1,Th2}, is indispensable for describing the features in the phase diagram associated with the A-like phase of liquid $^3$He in aerogel. Further, to examine effects of the local anisotropy, one needs to derive an expression of a disorder-induced term, corresponding to ${\cal S}_{2, {\rm dis}}$, in the GL action. In order to achieve these purposes {\it consistently}, we have chosen to work in the simplest Born approximation and its extention. To perform a more quantitative comparison between experimental and theoretical phase diagrams and obtain results on physical quantities such as the temperature dependence of $|\Delta|$ comparable with experimental data, the multiple scattering events and spatial variations of $|\Delta|$ need to be incorporated within a model of {\it anisotropic} and random scattering. 

\section{Free energy and gradient terms}

To evaluate free energy for various pairing states in the present disordered case, we will use the Gaussian variational method (GVM) . In this method, a {\it trial} Gaussian ansatz ${\overline {\cal S}}_g$ for the replicated action ${\overline {\cal S}}$ is first invoked. Then, the total free energy $F$ is evaluated as 
\begin{equation}\label{FGVM}
{\overline F} = {\overline F}_g + \frac{T}{n} \langle {\overline {\cal S}} - {\overline {\cal S}}_g \rangle_g, 
\end{equation}
where ${\overline F}_g$ is the free energy for ${\overline {\cal S}}_g$ divided by the number of replicas $n$, $\langle \, \, \, \rangle_g$ is the ensemble average on ${\overline {\cal S}}_g$, and the $n \to 0$ limit is taken at the end. The GVM has been satisfactorily applied in evaluating free energy of the random Ising-spin \cite{Dotsenko} and elastic systems \cite{GD}. 

To apply GVM to the present problem, we will first examine how to determine an appropriate trial action in our case with a couple of fields, the amplitude fluctuation $\delta |\Delta|=|\Delta|-|\Delta_{\rm MF}|$ and $a_{\mu,j}$ consisting of the symmetry variables. Since we are not interested in a negligibly narrow critical region in the close vicinity of $T_c$, we will assume, as usual, the two variables $\delta |\Delta|$ and $a_{\mu,j}$ to be separable in the {\it trial} Gaussian action. This assumption on the trial action greatly simplifies our analysis for eq.(\ref{FGVM}). In fact, the Gaussian approximation does {\it not} have to be assumed in the original action ${\overline {\cal S}}$ which appears only as its average in eq.(\ref{FGVM}). To clarify this point, let us rewrite the original gradient term 
\begin{eqnarray}\label{S2grad}
{\cal S}_{2,{\rm grad}} &=& T^{-1} \int d^3r \frac{1}{2} \biggl( K_1 \partial_i A^*_{\mu,i}({\bf r})  \partial_j A_{\mu,j}({\bf r}) \nonumber \\
&+& K_1 \partial_j A^*_{\mu,i}({\bf r})  \partial_i A_{\mu,j}({\bf r}) + K_2 \partial_j A^*_{\mu,i} \partial_j A_{\mu,i} \biggr).
\end{eqnarray}
included in ${\cal S}_2^{(0)}$. It is not difficult to see that the $K_1$ term in eq.(\ref{S2grad}) is rewritten as 
\begin{eqnarray}\label{gradsym}
&\int d^3r& \frac{K_1}{2T} \biggl[ |\Delta|^2 (\partial_i a_{\mu,i}^* \partial_j a_{\mu,j} + \partial_j a_{\mu,i}^* \partial_i a_{\mu,j} ) \nonumber \\
 &+& {\rm Re}(a_{\mu,i}^* a_{\mu,j}) (2 \partial_i|\Delta| \partial_j|\Delta| - \partial_i \partial_j |\Delta|^2 ) \biggr]
\end{eqnarray}
except surface terms. 
In the ABM or planar state, the presence of $a_{\mu,i}$ in the 2nd line of eq.(\ref{gradsym}) makes this term a nonGaussian form, because the factor ${\rm Re}(a^*_{\mu,i} a_{\mu,j})$ becomes $(\delta_{i,j}- {\bf l}_i {\bf l}_j)/2$ there, although it is merely $\delta_{i,j}/3$ in the BW or robust state from the outset. In the disordered ABM or planar state, however, the ${\bf l}$-vector has no orientational long range order \cite{Volovik96}, and hence, the random average of the factor ${\rm Re}(a^*_{\mu,i} a_{\mu,j})$ is merely $\delta_{i,j}/3$ irrespective of the correlation range of the ${\bf l}$-orientation. In this way, the original gradient term of ${\cal S}$, if applied to eq.(\ref{FGVM}), can be replaced by  
\begin{eqnarray}\label{S2gradav}
{\cal S}_{2,{\rm grad}} &\simeq& \frac{1}{T} \int d^3r \biggl[ {\tilde K} (\nabla |\Delta|)^2 + \frac{|\Delta|_{\rm MF}^2}{2} [ \, K_2 \partial_i a_{\mu,j} \partial_i a^*_{\mu,j} \nonumber \\
&+&  K_1 
(\partial_i a_{\mu,i} \partial_j a^*_{\mu,j} + \partial_j a_{\mu,i} \partial_i a^*_{\mu,j} ) \, ] \biggr]
\end{eqnarray} 
for {\it all} pairing states considered in this paper, where ${\tilde K} = (3K_2 + 2K_1)/6$. Here, according to the assumption of a slowly varying $a_{\mu,j}$ mentioned below eq.(\ref{sdis2}), the factor $|\Delta|^2$ was replaced by its uniform value $|\Delta_{\rm MF}|^2$ to be determined later. 

Then, in the total and averaged free energy ${\overline F} = {\overline F}_{\rm amp} + {\overline F}_{\rm sym}$, the $\delta |\Delta|$-part ${\overline F}_{\rm amp}$ and the $a_{\nu,j}$ - part ${\overline F}_{\rm sym}$ can be treated independently below:
\begin{eqnarray}
{\overline F}_{\rm amp} &=& -T \lim_{n \to +0} \frac{{\overline {Z^n}}_{\rm amp} - 1}{n}, \nonumber \\
{\overline F}_{\rm sym} &=& -T \lim_{n \to +0} \frac{{\overline {Z^n}}_{\rm sym} - 1}{n},
\end{eqnarray}
\begin{eqnarray}\label{partition}
{\overline {Z^n}}_{\rm amp} &=& {\rm Tr}_{\delta|\Delta|} \exp(-{\overline {\cal S}}_{\rm amp}), \nonumber \\ 
{\overline {Z^n}}_{\rm sym} &=& {\rm Tr}_{{\tilde A}_{\mu,i}} \exp(-{\overline {\cal S}}_{\rm sym}), 
\end{eqnarray}
where ${\tilde A}_{\mu,i} = |\Delta_{\rm MF}| a_{\mu,i}$. Since variations of $a_{\mu,i}$ are always accompanied by $\Delta_{\rm MF}$ in ${\cal S}_{2,{\rm grad}}$, the free energy correction due to the {\it purely} thermal fluctuation of symmetry variables is independent \cite{Kleinert} of $|\Delta_{\rm MF}|$ and thus, of the details of pairing states. Since such a free energy correction insensitive to $|\Delta_{\rm MF}|$ should take a common value to all $p$-wave pairing states
, this purely thermal correction will not be considered in $\langle {\overline {\cal S}} \rangle_g$ hereafter in examining a {\it relative} stability between different pairing states. 

According to treatments performed so far, the replicated action ${\overline {\cal S}}_{\rm amp}$ for the ABM state is given by 
\begin{eqnarray}\label{Samp0}
{\overline {\cal S}}_{\rm amp} &=& T^{-1} \sum_{a=1}^n \int d^3r \biggl[ \alpha |\Delta^{(a)}|^2 + {\tilde K} (\nabla |\Delta^{(a)}|)^2 \nonumber \\
&+& \beta_{\rm ABM} |\Delta^{(a)}|^4 
- \frac{3}{5} \delta \beta_{\rm d} \sum_{b=1}^n |\Delta^{(a)}|^2 |\Delta^{(b)}|^2 \biggr]. 
\end{eqnarray}
The corresponding expression for the BW (robust) state is given by replacing $\beta_{\rm ABM}$ and the factor $3/5$ in the second line by $\beta_{\rm BW}$ ($\beta_{\rm R}$) and $1/2$, respectively, while the corresponding one in the planar state follows from replacing $\beta_{\rm ABM}$ by $\beta_{\rm P}$. 

On the other hand, the replicated action, ${\overline {\cal S}}_{\rm sym} \equiv {\overline {\cal S}}_{\rm grad} + {\overline {\cal S}}_{{\rm dis}(2)}$, for the ABM and planar states is 
\begin{eqnarray}\label{Ssymn}
{\overline {\cal S}}_{\rm grad} \! &\simeq& \! \frac{|\Delta_{\rm MF}|^2}{2T} \! \int d^3r \sum_{a=1}^n \biggl[ 2 K_1 \partial_i a_{\mu,i} \partial_j a^*_{\mu,j} \nonumber \\
&+& K_2 \partial_i a_{\mu,j} \partial_i a^*_{\mu,j} \biggr], \nonumber \\ 
{\overline {\cal S}}_{{\rm dis}(2)} &\simeq& - \frac{3}{20 T} \delta \beta_{\rm d} |\Delta_{\rm MF}|^4 \sum_{b=1}^n ( \, ({\bf l}^{(a)}\cdot{\bf l}^{(b)})^2 - 1 \, ) 
\end{eqnarray}
if the field $a_{\mu,i}$ in the planar state is represented by eq.(C1) in Appendix C. 

Here, for later convenience, the gradient energy in the purely ABM pairing state will be expressed in the hydrodynamic representation \cite{VW,Cross} 
\begin{eqnarray}\label{gradAg}
{\overline {\cal S}}_{\rm grad} &=& {\overline {\cal S}}_{\rm Fr} + \frac{1}{2T} \int d^3r \sum_{a=1}^n \biggl[ \rho_0 \, M_{ij}^{(a)} (v^{(a)})_i \, (v^{(a)})_j \nonumber \\
&-& 2 b C \, {\bf v}^{(a)}\cdot{\bf L}^{(a)} + 2 C \, {\bf v}^{(a)}\cdot{\rm curl}  {\bf l}^{(a)}  \biggr], 
\end{eqnarray}
where 
\begin{equation}
{\bf v}_i = {\bf m}_j \nabla_i {\bf n}_j, \,\,\,\, {\bf L} = {\bf l} ({\bf l}\cdot{\rm curl}{\bf l}), \,\,\,\, 
M_{ij} = \delta_{i,j} - A \, {\bf l}_i {\bf l}_j, 
\end{equation} 
with positive constants $A$ and $b$, and ${\overline {\cal S}}_{\rm Fr}$ is the replicated Frank energy term 
\begin{eqnarray}
{\overline {\cal S}}_{\rm Fr} &=& \frac{1}{2T} \int d^3r \sum_{a=1}^n \biggl[ K_s ({\rm div}{\bf l}^{(a)})^2 + K_t \, ({\bf L}^{(a)})^2 \nonumber \\
&+& K_b ({\bf l}^{(a)}\cdot \nabla {\bf l}^{(a)})^2 \biggr]
\end{eqnarray}
in the terminology of the nematic liquid crystal, if the ${\bf l}$-vector is identified with the nematic director. 

On the other hand, for the the BW and robust states, the $\delta \beta_{\rm d}$ term of eq.(\ref{Ssymn}) is absent. Since, as mentioned earlier, the thermal fluctuation term of the symmetry variables is unnecessary for the present purpose, even $\langle {\overline {\cal S}}_{\rm grad} \rangle_g$ in eq.(\ref{FGVM}) may be neglected. Therefore, for the BW and robust states, we have no contribution of ${\overline F}_{\rm sym}$, and the total free energy ${\overline F}$ can be identified with ${\overline F}_{\rm amp}$. 

Now, let us turn to evaluating free energy of the disordered ABM state. The corresponding results for other pairing states will be commented on at the end of this section. First, to examine ${\overline F}_{\rm amp}$, it is convenient to rewrite eq.(\ref{Samp0}) in the form expressed in terms of a scalar order parameter $\phi({\bf r})$
\begin{eqnarray}\label{Sising}
{\overline {\cal S}}_{\rm ising} &=& \int d^3r \sum_{a,b} \biggl[ \delta_{a,b} \biggl( \frac{\tau_0}{2} (\phi^{(a)})^2 + \frac{1}{2}(\nabla \phi^{(a)})^2 
\nonumber \\ 
&+& \frac{g}{4} (\phi^{(a)})^4 \biggr) - \frac{u}{4} (\phi^{(a)} \phi^{(b)})^2 \biggr], 
\end{eqnarray}
which was studied within GVM in Ref.\cite{Dotsenko} as a continuum model of a random Ising spin system. Here, the scale transformation, $|\Delta|^2 ({\tilde K})^{3/2}/(T (N(0))^{1/2}) \to \phi^2/2$ and $[N(0)/{\tilde K}]^{1/2} {\bf r} \to {\bf r}$, was performed. Details of derivation of the free energy for the model (\ref{Sising}) is explained in Appendix B. By rewriting eq.(\ref{fampf}), the resulting ${\overline F}_{\rm amp}$ is found to take the form 
\begin{eqnarray}\label{FampA}
\frac{{\overline F}_{\rm amp}}{V} &=& - \frac{(N(0))^2}{4\beta_{\rm ABM}} \lambda_p^2 - \frac{T (N(0))^{3/2}}{2 \pi {\tilde K}^{3/2}} \frac{p_c}{2 \pi} |\lambda_p| 
\nonumber \\
&-& \frac{T (N(0))^{3/2}}{4 \pi^2 {\tilde K}^{3/2}} (3g-2u) \big(\frac{p_c}{2 \pi} \big)^2, 
\end{eqnarray}
where $V$ is the volume, 
\begin{equation}\label{lamA}
\lambda_p= \frac{\alpha}{N(0)} + (3g-2u)\frac{p_c}{2 \pi^2}, 
\end{equation}
and 
\begin{eqnarray}
g &=& \frac{T \beta_{\rm ABM}}{(N(0))^{1/2} {\tilde K}^{3/2}}, \nonumber \\
u &=& \frac{3 \delta \beta_{\rm d}}{5 \beta_{\rm ABM}} g. 
\end{eqnarray}
The dimensionless momentum cutoff $p_c/(2 \pi)$ will be assumed below to 
be unity. We note that the third term of eq.(\ref{FampA}) merely gives a negligibly small correction to the first and second ones in the relative difference between free energies of two different pairing states, since $g$ and $u$ are at most O($T^2/E_{\rm F}^2$). Depending on the disorder strength, this correction may be negligible compared with the contribution from ${\overline F}_{\rm sym}$ which will be examined below (Note that ${\overline F}_{\rm sym}$ is absent in the BW and robust states). 

In contrast to ${\overline F}_{\rm amp}$, it is not tractable to directly evaluate ${\overline F}_{\rm sym}$ in the ABM state. To evaluate ${\overline F}_{\rm sym}$ in a different manner, let us first start from examining free energy of the simpler model \cite{GD} 
\begin{eqnarray}\label{SGD}
{\overline {\cal S}}_{\rm XY} &=& \frac{1}{2} \sum_a \int d^3r \, [ \, {\tilde c} (\nabla \theta^{(a)})^2 \nonumber \\
&+& T^{-2} {\tilde W} \sum_b ( 1 - {\rm cos} [2(\theta^{(a)} - \theta^{(b)})] \, ) \, ].
\end{eqnarray}
Assuming a Gaussian trial action  
\begin{equation}
{\overline {\cal S}}_{\rm tr} = \frac{1}{2} \sum_{\bf q} \sum_{a,b} {\tilde {\cal G}}^{-1}_{ab}({\bf q}) \theta^{(a)}({\bf q}) \theta^{(b)}(-{\bf q}), 
\end{equation}
the corresponding averaged free energy ${\overline F}_{\rm XY}$ is given by 
\begin{eqnarray}\label{FGDGVM}
\frac{{\overline F}_{\rm XY}}{T V} &=& \frac{1}{2n} \biggl[ {\tilde c} \int_{\bf q} q^2 \, {\rm Tr} {\bf {\tilde {\cal G}}}({\bf q}) + \int_{\bf q} {\rm Tr} \, {\rm ln} \, [{\bf {\tilde {\cal G}}}^{-1}({\bf q})] \nonumber \\
&-& \frac{\tilde W}{T^2} (\sum_{a \neq b} \exp{(-2 B_{ab}(0))} + n)
\end{eqnarray}
except a constant term, where $B_{ab}(0) = \int_{\bf q} ({\tilde {\cal G}}_{aa}({\bf q}) + {\tilde {\cal G}}_{bb}({\bf q}) - 2 {\tilde {\cal G}}_{ab}({\bf q}) )$, and the $n \to 0$ limit is taken at the end. By following the procedures in Ref.\cite{GD}, the disorder dependent term of the first term of eq.(\ref{FGDGVM}) is 
given by 
\begin{equation}
\frac{1}{2} \int_{\bf q} \int_0^1 \frac{du}{u^2} \frac{[\sigma]_u}{{\tilde c} q^2 + [\sigma]_u},  
\end{equation}
while the integrand in its second term is expressed by 
\begin{equation}
{\rm Tr} \, {\rm ln} \, [{\bf {\tilde {\cal G}}}^{-1}({\bf q})] = n \biggl[ {\rm ln}({\tilde c} q^2) - \int_0^1 \frac{du}{u^2} {\rm ln}\biggl(\frac{[\sigma]_u}{{\tilde c} q^2} + 1 \biggr) \biggr].
\end{equation}
Details of the function $[\sigma]_u$ can be seen in Ref.\cite{GD}. Using the properties of $[\sigma]_u$ carrying the disorder strength ${\tilde W}$, the ${\bf q}$-integral of the second term of eq.(\ref{FGDGVM}) can be shown to be convergent. Then, it is found that, up to the lowest order in the disorder strength, the sum of the first and second terms in eq.(\ref{FGDGVM}) is disorder-independent. Therefore, 
the change of free energy density induced by the quenched disorder is, up to O(${\tilde W}$), given by the second line of eq.(\ref{FGDGVM}), i.e.,  
\begin{eqnarray}\label{FGDf}
\frac{{\overline F}_{\rm XY}({\tilde W}) - {\overline F}_{\rm XY}(0)}{V} &=& \frac{{\tilde W}}{2 T} \biggl(\exp{\biggl(-4 \int_{\bf q} \frac{1}{{\tilde c} q^2} \biggr)} - 1 \biggr) \nonumber \\
&\simeq& - \frac{p_c}{\pi^2} \frac{{\tilde W}}{{\tilde c} \, T} 
\end{eqnarray}
which is independent of $T$ because ${\tilde c}$ is an elastic constant divided by $T$. Note that eq.(\ref{FGDf}) is determined by the behavior at short scales of O($p_c^{-1}$), implying that the free energy is unaffected by the details of long distance behaviors \cite{comdipole}, i.e., the presence or absence of quasi LRO. In fact, reflecting the fact \cite{GD} that the elastic behavior at short scales is determined within the replica-symmetric approximation, the result (\ref{FGDf}) coincides with the corresponding one of the random-force model \cite{Larkin,IM} 
\begin{equation}\label{rfm}
{\cal S}_{\rm RF} = \int d^3r \biggl[ \frac{{\tilde c}}{2} (\nabla \theta)^2 + f({\bf r}) \theta({\bf r}) \biggr],
\end{equation}
where ${\overline f} = 0$, and 
\begin{equation}
{\overline {f({\bf r}) f({\bf r}')}} = 4 T^{-2} {\tilde W} \delta^{(3)}({\bf r} - {\bf r}').
\end{equation}
This action is equivalent to the Gaussian replicated action obtained from eq.(\ref{SGD}) with the replacement $1 - {\rm cos}(2(\theta^{(a)} - \theta^{(b)})) \to 2(\theta^{(a)} - \theta^{(b)})^2$. 

Based on this fact for the model (\ref{SGD}), we have evaluated ${\overline F}_{\rm sym}$ by, in the last term of ${\overline {\cal S}}_{\rm sym}$, keeping only the lowest (harmonic) order terms in Euler angles representing the ${\bf l}$-vector. Then, if using the representation 
\begin{equation}\label{lvector}
{\bf l} = {\hat z} \, {\rm cos}\theta_l + ({\hat x} \, {\rm cos}\phi_l + {\hat y} \, {\rm sin}\phi_l) \, {\rm sin}\theta_l, 
\end{equation}
one finds that the resulting last term of ${\overline {\cal S}}_{\rm sym}$ takes the form 
\begin{equation}\label{Sdisis}
{\overline {\cal S}}_{{\rm dis}(2)} \simeq \frac{3}{20 T} \delta\beta_{\rm d} |\Delta_{\rm MF}|^4 \sum_{a,b} \int d^3r (\theta_l^{(a)} - \theta_l^{(b)})^2 
\end{equation}
which depends only on $\theta_l^{(a)}$ and $\theta_l^{(b)}$. Further, the gradient energy, eq.(\ref{gradAg}), in the ABM state will be replaced, for simplicity, by its isotropized version 
\begin{eqnarray}\label{gradAis}
{\overline {{\cal S}_{\rm grad}^{({\rm iso})}}} &=& \frac{1}{2T} \int d^3r \sum_{a=1}^n \biggl[ {\rho}^{({\rm iso})} (v^{(a)})^2 \nonumber \\ 
&+& 2 C^{({\rm iso})} \, {\bf v}^{(a)} \cdot {\rm curl} {\bf l}^{(a)} \biggr] 
+ {\overline {\cal S}}_{\rm Fr}
\end{eqnarray}
corresponding to the limit of a Bose gas of molecules with the ABM pairing symmetry \cite{Volovikusp}, where 
${\rho}^{({\rm iso})}$ and ${C}^{({\rm iso})}$ are averaged coefficients which follow by replacing, e.g., ${\bf l}\cdot {\bf v} {\bf l}\cdot({\rm curl}{\bf l})$ in the original action by $\langle {\bf l}_i {\bf l}_j \rangle v_i ({\rm curl} {\bf l})_j$ and applying the absence of the ${\bf l}$-orientational LRO. The second term, proportional to ${\bf v} \cdot {\rm curl}{\bf l}$, will be neglected hereafter because it simply becomes a sum of purely surface terms after expressing it via the Euler angles. This easily follows from the fact that, in the representation (\ref{lvector}), ${\bf v} \cdot {\rm curl}{\bf l}$ is proportional to $(\nabla {\rm cos}(2\theta_l) \times \nabla \phi_l)_z - 2(\nabla {\rm cos}^2\theta_l \times \nabla {\rm sin}\phi_l)_x + 2(\nabla {\rm cos}^2\theta_l \times \nabla {\rm cos}\phi_l)_y$. 
Further, the remaining terms expressed in terms of the Euler angles, $\theta_l$ and $\phi_l$, will also be linearlized by using the absence of LRO. For instance, ${\rm sin}(2\theta_l) (\nabla \theta_l)^2$ will be replaced by $\langle {\rm sin}(2\theta_l) \rangle (\nabla \theta_l)^2$ which vanishes due to the absence of LRO. The resulting expression is Gaussian in $\nabla \phi_l$ and $\nabla \theta_l$, and there are no cross terms like $\nabla \theta_l \nabla \phi_l$ there. In fact, the Gaussian term in $\nabla \theta_l$, which is the relevant one for the present purpose (see eq.(\ref{Sdisis})), results only from ${\overline {\cal S}}_{\rm Fr}$. In this way, the relevant gradient energy term in ${\overline {\cal S}}_{\rm sym}$ becomes 
\begin{equation}\label{Ssymis}
\frac{5}{18} K_b \sum_{a=1}^n \int d^3r (\nabla \theta_l^{(a)})^2
\end{equation}
in the weak coupling approximation where $K_b=3 K_s = 3 K_t = 3|\Delta_{\rm MF}|^2 (K_1+K_2)/4$. The coefficients $K_b$, $K_s$, and $K_t$ including the strong coupling corrections are given, up to the lowest order in $(T_{c} - T)/T_{c}$, by their weak coupling expressions divided by the mass enhancement factor \cite{Fetter}, if $\delta \beta_j^{({\rm sc})}$ and $\delta {\tilde \beta}_j^{({\rm sc})}$ are properly incorporated in $\beta_j$ appearing in $|\Delta_{\rm MF}|$. Thus, eq.(\ref{Ssymis}) is expected to be applicable even at higher pressures as far as pressure dependences of $\xi_0$ and $N(0)$ are incorporated through their experimental data. 
The remaining $\phi_l$-dependent terms are purely thermal fluctuation contributions unrelated to the quenched disorder and hence, may be neglected hereafter to derive the $\delta \beta_{\rm d}$-dependent correction to the free energy. Then, eq.(\ref{Ssymis}) accompanied by eq.(\ref{Sdisis}) is of the same form as the random-force model, eq.(\ref{rfm}), if $3 T \delta \beta_{\rm d} |\Delta_{\rm MF}|^4/20$ is identified with ${\tilde W}$, and hence, the resulting disorder contribution to ${\overline F}_{\rm sym}$ is given by 
\begin{equation}\label{DFsymA}
\frac{{\overline F}_{\rm sym}(\delta \beta_{\rm d}) - {\overline F}_{\rm sym}(0)}{V} = \frac{- 9 T N(0) |\lambda_p| \delta \beta_{\rm d}}{25 \pi \beta_{\rm ABM}(K_1 + K_2) \xi_0} \, \frac{p_c}{2 \pi}.
\end{equation}

We are now at the stage of discussing stability of the pairing states and the resulting phase diagram of superfluid $^3$He in globally isotropic aerogels. To perform the remaining task, we need just the free energy expressions, eqs.(\ref{FampA}) and (\ref{DFsymA}), and information on the SC effects in each state (see Appendix A and Ref.\cite{AI07}). First, 
judging from the data of SC parameters, there is no possibility that the ABM state is replaced by the robust phase \cite{Fomin}. The contributions from the $\delta \beta_{\rm d}$ term to the free energy definitely show that this term favors the anisotropic ABM and planar states. Although the disorder effect on the SC parameters may suggest a small gain of the condensation energy in the robust state, it is quite difficult to, in the weak disorder regime, find such a situation that the robust state is realized due to an enhanced disorder. Rather, it is more reasonable to examine the planar state as a candidate, other than the ABM one, of the A-like phase. However, since inevitably $\beta_{\rm P} > \beta_{\rm ABM}$, ${\overline F}_{\rm amp}$ in the planar state is higher than that of the ABM state. In addition to this, the planar state is not favored even through ${\overline F}_{\rm sym}$: 
As shown in Appendix C, the gradient energy in ${\overline {\cal S}}_{\rm sym}$ of the planar state is 2.4 times bigger than that of the ABM case. Since the expression for the planar state corresponding to eq.(\ref{DFsymA}) is also inversely proportional to the magnitude of the gradient energy, the free energy gain in the planar state due to the random symmetry variables is much smaller than that of the ABM state. By taking account of these results on ${\overline F}$ altogether, we conclude that even the planar pairing state cannot become stable as the A-like phase in the GL region in $^3$He in aerogels. 

\begin{figure}[t]
\scalebox{0.5}[0.5]{\includegraphics{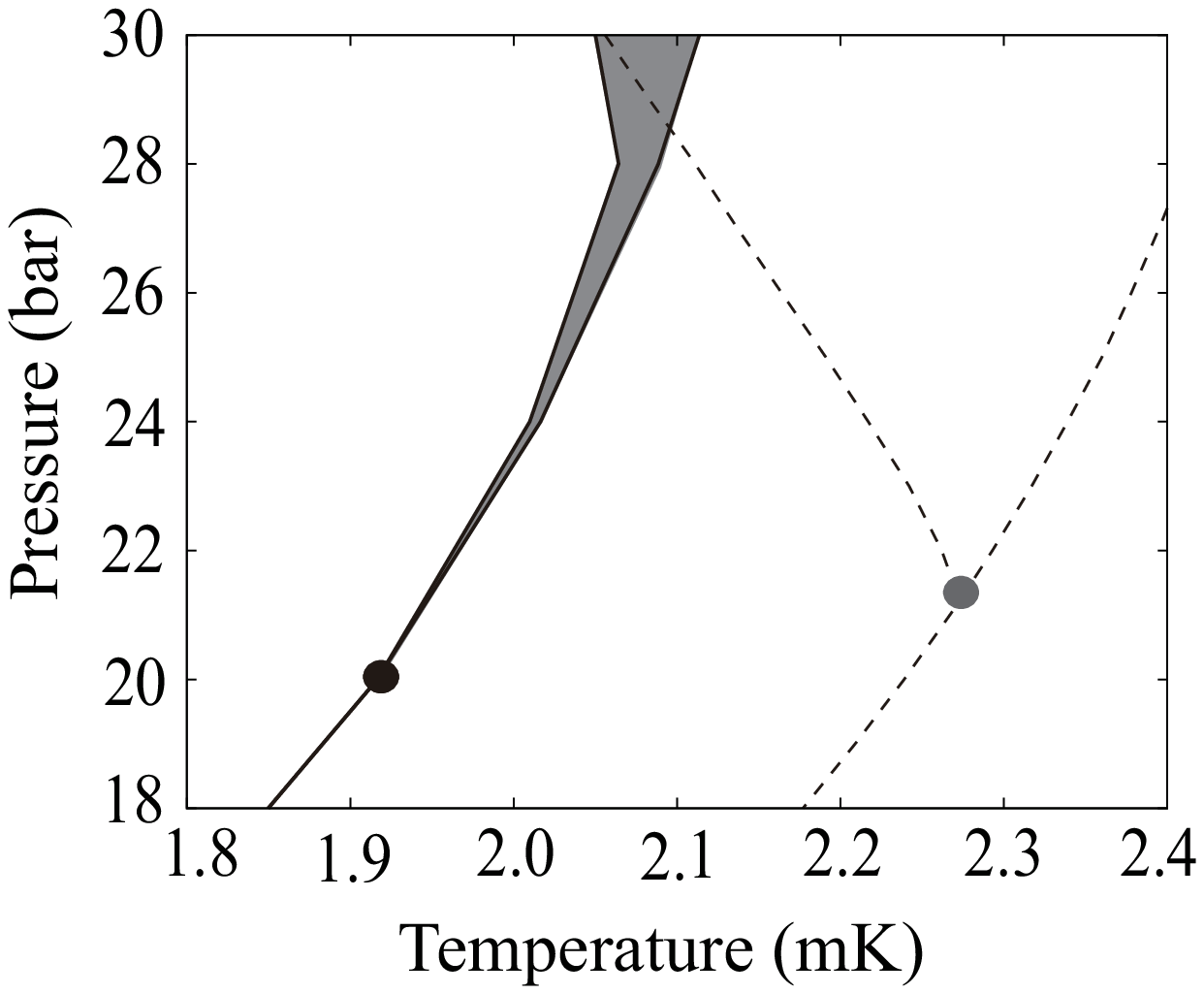}}
\caption{Example of calculated $P$-$T$ phase diagrams obtained based on the present free energy analysis. The solid curves denote the transition curves in the disordered case specified by the parameter values $(2 \pi \tau T_{c0})^{-1}=0.058$ for $P=20$(bar) and $\gamma=21$, while the dashed ones are the corresponding bulk transition curves. The hatched region indicates the A-like phase region. The $T_{c0}(P)$ and $E_{\rm F}(P)$ data are taken from Table 4.1 of Ref.\cite{VW} and Table VI of Ref.\cite{RMP75}, respectively, while the analysis on the SC correction entirely follows the phenomenological method in Ref.\cite{AI07}. }
\end{figure}

In Fig.3, a typical example of pressure v.s. temperature phase diagrams we obtain is shown. There, eqs.(\ref{FampA}) and (\ref{DFsymA}) were used based on data of pressure dependences of $E_{\rm F}$ and $T_{c0}$. The disorder-induced reduction of PCP indicated as a solid circle is a consequence of the large $\gamma$-value used here. The parameter $\gamma$ was defined in eq.(\ref{locani}) as a measure of the strength of the local anisotropy in the scattering events. The fact that the A-like phase is limited to such a narrow temperature range is a combined effect of $T_c$-reduction and the SC effect \cite{AI07} shrinking with increasing disorder. 

\section{Quasi long range orientational order in A-like phase}

In the preceding section, a typical phase diagram following from evaluation of free energy was shown in Fig.3. However, it is important to note that, at the present stage, the transition curves in the figure merely imply changes of the pairing states. As noted in sec.I, if the A-like phase of $^3$He in aerogel is in a disordered ABM pairing state, the genuine long range order of ${\bf l}$-orientation is absent in the A-like phase \cite{Volovik96}. It is crucial to clarify whether such a 3D phase with no genuine long range superfluid order may show superfluidity (see sec.I). In this section, we address this possibility at weak disorder where the singular topological defects are not excited via the disorder. This issue is highly nontrivial because, in the globally isotropic case, the nonsingular vortices \cite{MH} or vortex-skyrmions \cite{Volovik07} may appear as a result of a disorder-induced ${\bf l}$-texture and thus, may destroy superfluidity. It is shown below that a one-loop renormalization of the gradient energy terms accompanying the functional RG treatment \cite{Feldman,DFisher,Nattermann} yields only a stable fixed point at which the vortex-skyrmions are irrelevant. This implies that the A-like phase at weak disorder may have {\it quasi} long range superfluid order {\it and} superfluidity. 

To examine long distance behaviors of the symmetry variables of the disordered ABM pairing state, we examine 
${\overline F}_{\rm sym}$ again. For the sake of the ensuing analysis, however, the kinetic part of ${\cal S}_{\rm grad}$ expressed in the form of eq.(\ref{gradAg}) will be rewritten in the form 
\begin{eqnarray}\label{Skin}
{\overline {\cal S}}_{\rm grad} - {\overline {\cal S}}_{\rm Fr} &=& \frac{1}{2T} \int_{\bf r} \sum_{a=1}^n \biggl( \, \rho_0 M_{ij}^{(a)} {\tilde {\bf v}}_i^{(a)} {\tilde {\bf v}}_j^{(a)} \nonumber \\
&-& \frac{b^2 C^2}{\rho_0 (1-A)} ({\bf L}^{(a)})^2 \, \biggr),
\end{eqnarray}
where $\int_{\bf r}$ denotes $\int d^3r$, and ${\tilde {\bf v}} = {\bf v} - b C \, {\bf L}/[ \rho_0 (1-A) ]$. Next, eliminating the longitudinal component of ${\bf v}$, ${\tilde {\bf v}}$ in eq.(\ref{Skin}) is replaced by its transverse component $\int d^3r' (\nabla' \times (\nabla' \times {\tilde {\bf v}}'))/(4 \pi \, |{\bf r} - {\bf r}'|)$, and eq.(\ref{Skin}) can be replaced by 
\begin{eqnarray}\label{Skin2}
{\overline {\cal S}}_{\rm grad} &-& {\overline {\cal S}}_{\rm Fr} = \frac{\rho_0}{32 \pi^2 T} \int_{\bf r} \int_{{\bf r}'} \int_{{\bf r}_1} \sum_{a=1}^n (\nabla \times (\nabla \times {\tilde {\bf v}^{(a)}({\bf r})}))_i \nonumber \\ 
&\times& M_{ij}^{(a)}({\bf r_1}) w({\bf r}-{\bf r}_1) w({\bf r}' - {\bf r}_1) (\nabla' \times (\nabla' \times {\tilde {\bf v}^{(a)}({\bf r}')}))_j 
\nonumber \\
&-& \int_{\bf r} \sum_{a=1}^n \frac{b^2 C^2}{ 2 T \rho_0 (1-A)} 
({\bf L}^{(a)})^2({\bf r}), 
\end{eqnarray}
where $w({\bf r}) = |{\bf r}|^{-1}$, and ${\bf v}'$ denotes ${\bf v}({\bf r}')$. Further, using $\nabla^2 w({\bf r}) = - 4 \pi \delta^{(3)}({\bf r})$ 
and rewriting the terms in eq.(\ref{Skin2}) consisting only of $\nabla \times {\bf L}^{(a)}$, we obtain the following action of the nonlocal gradient energy appropriate for the ensuing RG analysis 
\begin{widetext}
\begin{eqnarray}\label{startgrad}
{\overline {\cal S}}_{\rm grad} &=& {\tilde {\cal S}}_{\rm Fr} + \frac{1}{{\tilde T}} \int_{\bf r} \int_{{\bf r}'} w({\bf r} - {\bf r}') \sum_{a=1}^n \biggl[ \, \rho \, {\bf \Omega}^{(a)}({\bf r}) \cdot {\bf \Omega}^{(a)}({\bf r}') - \frac{\rho_1}{2} [ {\bf \Omega}^{(a)}({\bf r})\cdot(\nabla' \times {\bf L}^{(a)}({\bf r}')) +  {\bf \Omega}^{(a)}({\bf r}')\cdot(\nabla \times {\bf L}^{(a)}({\bf r}))] \nonumber \\
&-& \frac{{\overline \rho}_1}{2} [ {\bf \Omega}^{(a)}({\bf r})\cdot(\nabla' \times (\nabla' \times {\bf l}'^{(a)})) + {\bf \Omega}^{(a)}({\bf r}')\cdot(\nabla \times (\nabla \times {\bf l}^{(a)})) ] - \frac{\rho_2}{2} {\rm div}{\bf L}^{(a)}({\bf r}) \, {\rm div}'{\bf L}^{(a)}({\bf r}') \biggr] \nonumber \\ 
&-& \int_{{\bf r}_1} \sum_{a=1}^n \frac{{\bf l}^{(a)}_i({\bf r}_1) {\bf l}^{(a)}_j({\bf r}_1)}{2 {\tilde T}}  \int_{{\bf r}'} \int_{\bf r} [ \, \sigma \, ({\bf \Omega}^{(a)}({\bf r}) \times \nabla)_i ({\bf \Omega}^{(a)}({\bf r}') \times \nabla')_j - \sigma_1 \, [ ((\nabla \times {\bf L}^{(a)}({\bf r})) \times \nabla)_i ({\bf \Omega}^{(a)}({\bf r}') \times \nabla')_j \nonumber \\ 
&+&  ((\nabla' \times {\bf L}^{(a)}({\bf r}')) \times \nabla')_i ({\bf \Omega}^{(a)}({\bf r}) \times \nabla)_j] + \sigma_2 \, {\rm div} {\bf L}^{(a)}({\bf r}) \nabla_i \, {\rm div}' {\bf L}^{(a)}({\bf r}') \nabla'_j ] \, w({\bf r}-{\bf r}_1) w({\bf r}' - {\bf r}_1),
\end{eqnarray}
\end{widetext}
where 
\begin{equation}
{\bf \Omega}_i({\bf r}) = (\nabla \times {\bf v}({\bf r}))_i = \varepsilon_{ijk} {\bf l}\cdot(\nabla_j {\bf l} \times \nabla_k {\bf l})
\end{equation}
is the Mermin-Ho relation \cite{MH} in the ABM state, $\varepsilon_{ijk}$ is the antisymmetric tensor, and the following redefinition of the Frank energy term 
\begin{eqnarray}
{\overline {\cal S}}_{\rm Fr} &+& \frac{A}{2 T} \biggl(\frac{b^2 C^2}{\rho_0 (1 - A)^2} - \rho_0 \biggr) \int_{\bf r} \sum_{a=1}^n ({\bf L}^{(a)})^2 \equiv {\tilde {\cal S}}_{\rm Fr} \nonumber \\
&=& \frac{1}{2 {\tilde T}} \int_{\bf r} \sum_{a=1}^n ( \, \partial_\mu {\bf l}_\nu^{(a)} \partial_\mu {\bf l}_\nu^{(a)} + \lambda_2 ({\rm div} {\bf l}^{(a)})^2 \nonumber \\
&+& \lambda_3 (({\bf l}^{(a)}\cdot\nabla) {\bf l}^{(a)})^2 \, ) 
\end{eqnarray}
has been done. Further, the relations $\partial_\mu {\bf l}_\nu \partial_\mu {\bf l}_\nu = ({\rm div}{\bf l})^2 + L^2 + (({\bf l}\cdot\nabla) {\bf l})^2$ and ${\bf l}^2=1$ were used. Note that the coefficient $K_t$ of the twist deformation term was absorbed into $T$ to define ${\tilde T}$. The ${\overline \rho}_1$ term, which is absent in the bare action, has been included because it is generated via renormalization. The {\it bare} values of the seven coefficients except ${\overline \rho}_1$ in eq.(\ref{startgrad}) are positive, although their detailed expressions are not necessary in our analysis given below. In fact, it will be assumed that, through the dipole energy term, the ${\bf d}$-vector $d_\mu$ (see eq.(\ref{abmtext})) is locked into ${\bf l}$ at large scales of interest so that the gradient term on $d_\mu$ may be absorbed into the Frank energy. Nevertheless, we have the stability conditions 
\begin{equation}\label{stabineq}
\lambda_2 + 1 > 0, \, \, \, \, \, \lambda_3 + 1 > 0.
\end{equation}
The goal in this section is to find an action at a stable disorder-induced fixed point by examining the scaling of the coefficients. 

Following Ref.\cite{Feldman} in which a functional RG analysis was performed for ${\tilde {\cal S}}_{\rm Fr}$, let us focus on ${\tilde T} \to 0$ limit, in which thermal fluctuation effects arising from higher order terms in ${\tilde T}$ are neglected, and determine the form of ${\overline {\cal S}}_{\rm grad}$ at the stable fixed point. To perform this, the disorder energy term will be 
generalized to 
\begin{equation}
{\overline {\cal S}}_{\rm dis} = - \frac{1}{{\tilde T}^2} \int_{\bf r} \sum_{a,b=1}^n R({\bf l}^{(a)}({\bf r})\cdot{\bf l}^{(b)}({\bf r})). 
\end{equation}
In the functional RG analysis based on $\varepsilon=4-d$-expansion, a stable disorder-induced fixed point is determined by $R(z)$ of O($\varepsilon$) in magnitude and the fixed point values of the coefficients in ${\overline {\cal S}}_{\rm grad}$. To perform the one-loop renormalization of ${\overline {\cal S}}_{\rm grad}$, the ${\bf l}$-vector with the momentum ${\bf q}$ of O($1$) in magnitude will be expressed in terms of the "transverse" variables $\phi_j$ in the momentum-shell ($e^{-l} < |{\bf q}| < 1$) as \cite{Polyakov} ${\bf l}({\bf r})= {\overline {\bf l}}({\bf r}) \sqrt{1 - \sum_{j=1,2} (\phi^{(j)}({\bf r}))^2} \, + \, \sum_{j=1,2} \phi^{(j)}({\bf r}) {\bf e}^{(j)}({\bf r})$, where ${\bf e}^{(j)}\cdot{\tilde {\bf l}}={\bf e}^{(1)}\cdot{\bf e}^{(2)} = 0$, and the disorder function $R(z)$ will be expanded in powers of $\phi^{(j)}$. Further, when examining a renormalized ${\overline {\cal S}}_{\rm grad}$ up to one loop order, the replica-index dependences of ${\overline {\bf l}}$ and ${\bf e}^{(j)}$ may be neglected \cite{Feldman}. Then, we only have to examine the correction $-\langle \delta {\overline {\cal S}}_{\rm grad} \, \delta {\overline {\cal S}}_{\rm dis} \rangle$ to ${\overline {\cal S}}_{\rm grad}$ in ${\tilde T} \to 0$ limit, where $\delta {\overline {\cal S}}_{\rm grad}$ is the second order correction in $\phi^{(j)}$ to ${\overline {\cal S}}_{\rm grad}$, and 
\begin{equation}
\delta {\overline {\cal S}}_{\rm dis} = - \frac{1}{{\tilde T}^2} \int_{\tilde {\bf r}} R^{(1)}(1) \sum_{a=1}^n \sum_{j=1,2} ((\phi^{(j)})^{(a)}({\tilde {\bf r}}))^2, 
\end{equation}
where $R^{(1)}(1) = dR(z)/dz|_{z=1}$, and a term which vanishes in $n \to 0$ limit was neglected \cite{Feldman}. 

To illustrate the one-loop renormalization procedure, let us first focus on the isotropic limit with $A=b=0$ in which the original gradient energy is given by eq.(\ref{gradAis}). Alternatively, one may start from ${\tilde {\cal S}}_{\rm Fr} + (\rho/{\tilde T}) \int_{\bf r} \int_{{\bf r}'} w({\bf r} - {\bf r}') \sum_a {\bf \Omega}^{(a)}({\bf r})\cdot{\bf \Omega}^{(a)}({\bf r}')$ in place of eq.(\ref{startgrad}). For simplicity, the replica index $a$ and its summation will be omitted hereafter. 
By noting that the ${\bf v}\cdot{\rm curl}{\bf l}$ term has no bulk contribution and thus, is negligible, we find 
\begin{widetext}
\begin{eqnarray}
\delta {\overline {\cal S}}_{\rm grad}^{({\rm iso})} &=& \frac{\rho}{\tilde T} \int_{\bf r} \int_{{\bf r}'} w({\bf r} - {\bf r}') \biggl[ \, -\frac{3}{2} \sum_j ((\phi^{(j)})^2 + ((\phi^{(j)})')^2) {\bf \Omega}\cdot{\bf \Omega}' 
- 4 \sum_{j,k} \phi^{(j)} \phi^{(k)} {\bf e}^{(j)}_\rho \partial_\mu {\overline {\bf l}}_\rho {\bf e}^{(k)}_\lambda ({\overline {\bf l}} \times \partial_\nu {\overline {\bf l}})_\lambda \, \varepsilon_{\mu \nu \alpha} \Omega'_\alpha \nonumber \\
&+& 4\sum_{j,k} (\partial_\mu \phi^{(j)} {\overline {\bf l}}\cdot({\bf e}^{(j)} \times \partial_\nu {\overline {\bf l}}) - \partial_\nu \phi^{(j)} {\overline {\bf l}}\cdot({\bf e}^{(j)} \times \partial_\mu {\overline {\bf l}})) \, \partial'_\mu (\phi^{(k)})' {\overline {\bf l}'}\cdot(({\bf e}^{(k)})' \times \partial_\nu {\overline {\bf l}'}) \biggr] \nonumber \\
&+& \frac{1}{2{\tilde T}} \int_{\bf r} \sum_{j,k} [ {\bf e}^{(i)}_\rho \partial_\mu {\overline {\bf l}}_\rho {\bf e}^{(k)}_\lambda \partial_\mu {\overline {\bf l}}_\lambda - \partial_\mu {\overline {\bf l}}_\nu \partial_\mu {\overline {\bf l}}_\nu \delta_{j,k} + \lambda_2 [ ({\overline {\bf l}}\cdot\nabla){\overline {\bf l}}_\rho ({\overline {\bf l}}\cdot\nabla) {\overline {\bf l}}_\lambda {\bf e}^{(j)}_\rho {\bf e}^{(k)}_\lambda 
- ({\rm div} {\overline {\bf l}})^2 \delta_{j,k} ] \nonumber \\
&+& \lambda_3 [({\bf e}^{(j)}\cdot\nabla) {\overline {\bf l}}_\rho ({\bf e}^{(k)}\cdot\nabla) {\overline {\bf l}}_\rho + {\bf e}^{(j)}_\rho ({\overline {\bf l}}\cdot\nabla) {\overline {\bf l}}_\rho {\bf e}^{(k)}_\lambda ({\overline {\bf l}}\cdot\nabla){\overline {\bf l}}_\lambda - 2 (({\overline {\bf l}}\cdot\nabla) {\overline {\bf l}})^2 \delta_{j,k} ] \, ] \phi^{(j)} \phi^{(k)},
\end{eqnarray}
\end{widetext}
where the remaining terms harmonic in $\phi$ 
\begin{eqnarray}\label{nonintnem} 
\Delta {\tilde {\cal S}}_{\rm Fr} &=& \frac{1}{2{\tilde T}} \int_{\bf r} \sum_{j.k}[ \delta_{j.k} [(\nabla \phi^{(j)})^2 + \lambda_3 (({\overline {\bf l}}\cdot\nabla) \phi^{(j)})^2] \nonumber \\
&+& \lambda_2 (({\bf e}^{(j)}\cdot\nabla) \phi^{(j)} ({\bf e}^{(k)}\cdot\nabla) \phi^{(k)} ]  
\end{eqnarray}
can be identified with the "noninteracting" action for the 
$\phi^{(j)}$-fields. 

In evaluating $-\langle \delta {\overline {\cal S}}_{\rm grad} \delta {\overline {\cal S}}_{\rm dis} \rangle$, we encounter the following expressions in the momentum-shell 
\begin{eqnarray}
I({\bf r}) &=& \sum_{i,j,k} \int_{\tilde {\bf r}} \langle (\phi^{(i)}({\tilde {\bf r}}))^2 \phi^{(j)}({\bf r}) \phi^{(k)}({\bf r}) \rangle A_{jk}({\bf r}), \nonumber \\
I_{\mu,\nu}({\bf r}) &=& \sum_{i,j,k} \int_{\tilde {\bf r}} \int_{{\bf r}'} w({\bf r}-{\bf r}') \partial_\mu \partial'_\nu \langle (\phi^{(i)}({\tilde {\bf r}}))^2 \phi^{(j)}({\bf r}) \phi^{(k)}({\bf r}') \rangle \nonumber \\
&\times& B_{jk}({\bf r}; {\bf r}'-{\bf r}).
\end{eqnarray}
After the trivial integration in the momentum-shell, we easily obtain $I = {\tilde T}^2 \sum_i A_{ii}({\bf r}) J(\lambda_2, \lambda_3) (1 - e^{-l})$, where $1-e^{-l}$ is the thickness of the momentum-shell. Here, the $\lambda_2$ and $\lambda_3$ dependences of $J$ arise from the dependence of the "noninteracting" action, eq.(\ref{nonintnem}), on these coefficients. In all terms in the one-loop renormalization, however, the result of integration in the momentum-shell is expressed by the quantity $J(\lambda_2, \lambda_3)$, and its dependence on $\lambda_2$ and $\lambda_3$ is found not to affect the resulting fixed points and the linear stability around them. Thus, to simplify the ensuing expressions, the dependence of $J$ on $\lambda_2$ and $\lambda_3$ will be omitted hereafter. Then, using $\nabla^2 w({\bf r})= - 4\pi \delta^{(3)}({\bf r})$, we find $I_{\mu,\nu}({\bf r}) = 4 \pi {\tilde T}^2 J_0 \delta_{\mu,\nu} \sum_i B_{ii}({\bf r};0) (1 - e^{-l})/3$, where $J_0=J(0,0)$. 

Therefore, using the relations $\sum_j {\bf e}^{(j)}_\rho {\bf e}^{(j)}_\lambda = \delta_{\rho, \lambda} - {\overline {\bf l}}_\rho {\overline {\bf l}}_\lambda$ and $\partial_\mu {\bf e}^{(j)} \simeq - [{\bf e}^{(j)}_\lambda \partial_\mu {\overline {\bf l}}_\lambda] {\overline {\bf l}}$ \cite{Feldman}, we have 
\begin{widetext}
\begin{eqnarray}
-\langle \delta {\overline {\cal S}}_{\rm grad} \delta {\overline {\cal S}}_{\rm dis} \rangle &=& - (1 - e^{-l}) \frac{R^{(1)}(1) J_0}{\tilde T} \biggl[ 2 \rho \int_{\bf r} \int_{{\bf r}'} w({\bf r} - {\bf r}') {\overline \Omega}({\bf r})\cdot{\overline \Omega}({\bf r}') - 16 \pi \biggl(1 - \frac{1}{d} \biggr) \int_{\bf r} \partial_\mu {\overline {\bf l}}_\nu \partial_\mu {\overline {\bf l}}_\nu \nonumber \\
&+& \frac{1}{2} \int_{\bf r} [ (1 - \lambda_3)  \partial_\mu {\overline {\bf l}}_\nu \partial_\mu {\overline {\bf l}}_\nu + 2 \lambda_2 ({\rm div} {\overline {\bf l}})^2 + (4 \lambda_3 - \lambda_2) (({\overline {\bf l}}\cdot\nabla){\overline {\bf l}})^2 ] \biggr]. 
\end{eqnarray}
\end{widetext}
Taking account of the rescaling factor $e^{l(d-2)}$ of ${\tilde T}$ \cite{com}, we obtain the following recursion equations 
\begin{eqnarray}
\frac{d}{dl}{\tilde T}^{-1} &=& {\tilde T}^{-1} ( 2 -\varepsilon - J_0 R^{(1)}(1) (1 - \lambda_3 - {\hat \rho})), \nonumber \\
\frac{d \lambda_2}{dl} &=& - J_0 R^{(1)}(1) \lambda_2( \, 1 + \lambda_3 + {\hat \rho} \, ), \nonumber \\
\frac{d \lambda_3}{dl} &=& - J_0 R^{(1)}(1) ( \, 3 \lambda_3 - \lambda_2 + \lambda_3(\lambda_3 + {\hat \rho}) \, ), \nonumber \\
\frac{d {\hat \rho}}{dl} &=& - J_0 R^{(1)}(1) \, {\hat \rho}(\, 1 + \lambda_3 + {\hat \rho} \, ),
\end{eqnarray} 
where ${\hat \rho}=32 \pi \rho (1 - 1/d)$. The first equation simply ensures that, within the present analysis, we stay in the parameter space at zero temperature with no thermal fluctuation. Under the stability condition $1+\lambda_3 > 0$, the following two fixed 
points 
\begin{eqnarray}
({\rm i}) \,\,\, \lambda_2^* &=& \lambda_3^*={\hat \rho}^*=0, \nonumber \\
({\rm ii}) \, \, \, \lambda_2^* &=& \lambda_3^*/2, \,\,\, {\hat \rho}^* = -1-\lambda_3^* < 0
\end{eqnarray}
are found. 
The case (i) expresses the nematic fixed point \cite{Feldman} with no vortex-skyrmions which is easily shown through a linear stability analysis to be a stable fixed point. On the other hand, the case (ii) expressing a fixed line has a negative value of ${\hat \rho}$. However, this negative value does {\it not} imply a proliferation of the vortex-skyrmions induced by disorder, because this finite ${\hat \rho}^*$ is independent of the recursion equation of the disorder function $R(z)$. This physically unaccepted ${\hat \rho}$-value certainly indicates that this fixed line is a unphysical one. In this way, within the model of the isotropic gradient energy, the quasi long-range order of the orbital orientation, controlled by the nematic fixed point \cite{Feldman}, is found to be stable against the vortex-skyrmions. 

To verify whether the above result is affected by the "orbital anisotropy" leading to the finite $A$ and $b$, the same analysis as in the isotropic case will be applied to the full action (\ref{startgrad}). Through lengthy but straightforward calculations, we find that the one-loop recursion equations of the coefficients in eq.(\ref{startgrad}) are given by 
\begin{widetext}
\begin{eqnarray}
\frac{d {\tilde T}^{-1}}{dl} &=& {\tilde T}^{-1}( 2 - \varepsilon - J_0 R^{(1)}(1)(1 - \lambda_3 - {\hat \rho}+ {\hat \rho}_1 + {\hat \rho}_2 + 2 {\hat \sigma}_2), \nonumber \\
\frac{d \lambda_2}{dl} &=& - J_0 R^{(1)}(1) [ \, \lambda_2( \, 1 + \lambda_3 + {\hat \rho} - {\hat \rho}_1 - {\hat \rho}_2 - 2 {\hat \sigma}_2) - {\hat \rho}_1 \, ], \nonumber \\
\frac{d \lambda_3}{dl} &=& - J_0 R^{(1)}(1) ( \, 3 \lambda_3 - \lambda_2 + \lambda_3(\lambda_3 + {\hat \rho} - {\hat \rho}_1 - {\hat \rho}_2 - 2 {\hat \sigma}_2  ) - 2 {\hat \rho}_1 - {\hat \rho}_2 + 3 {\hat \sigma} - 8 {\hat \sigma}_1 - {\hat \sigma}_2 \, ), \nonumber \\
\frac{d {\hat \rho}}{dl} &=& - J_0 R^{(1)}(1) [ \, {\hat \rho}(\, 1 + \lambda_3 + {\hat \rho}  - {\hat \rho}_1 - {\hat \rho}_2 - 2 {\hat \sigma}_2 \, ) + {\hat \sigma} \, ], \nonumber \\
\frac{d {\hat \rho}_1}{dl} &=& - J_0 R^{(1)}(1) [ \, {\hat \rho}_1(\, 3 + \lambda_3 + {\hat \rho}  - {\hat \rho}_1 - {\hat \rho}_2 - 2 {\hat \sigma}_2 \, ) - 12 {\hat \sigma} \, ], \nonumber \\
\frac{d {\hat \rho}_2}{dl} &=& - J_0 R^{(1)}(1) [ \, {\hat \rho}_2 (\, 5 + \lambda_3 + {\hat \rho}  - {\hat \rho}_1 - {\hat \rho}_2 - 2 {\hat \sigma}_2 \, ) + {\hat \sigma}_1 + 10 {\hat \sigma}_2 \, ], \nonumber \\
\frac{d {\hat {\overline \rho}}_1}{dl} &=& - J_0 R^{(1)}(1) [ \, {\hat {\overline \rho}}_1 (\, 1 + \lambda_3 + {\hat \rho}  - {\hat \rho}_1 - {\hat \rho}_2 - 2 {\hat \sigma}_2 \, ) - {\hat \rho}_1 - 6 {\hat \sigma} \, ], \nonumber \\
\frac{d {\hat \sigma}}{dl} &=& - J_0 R^{(1)}(1) \, {\hat \sigma} (\, 4 + \lambda_3 + {\hat \rho}  - {\hat \rho}_1 - {\hat \rho}_2 - 2 {\hat \sigma}_2 \, ), \nonumber \\
\frac{d {\hat \sigma}_1}{dl} &=& - J_0 R^{(1)}(1) \, {\hat \sigma}_1 (\, 4 + \lambda_3 + {\hat \rho}  - {\hat \rho}_1 - {\hat \rho}_2 - 2 {\hat \sigma}_2 \, ), \nonumber \\ 
\frac{d {\hat \sigma}_2}{dl} &=& - J_0 R^{(1)}(1) \, {\hat \sigma}_2 (\, 8 + \lambda_3 + {\hat \rho}  - {\hat \rho}_1 - {\hat \rho}_2 
- 2 {\hat \sigma}_2 \, ), 
\end{eqnarray}
\end{widetext}
where ${\hat \rho}_1 = 32 \pi \rho_1$, ${\hat \rho}_2 = 8 \pi \rho_2/d$, ${\hat {\overline \rho}}_1 = 32 \pi {\overline \rho_1}$, ${\hat \sigma}=128 \pi^2 \sigma (1 - 1/d)$, ${\hat \sigma}_1=8 \pi^2 \sigma_1 (1 - 1/d)$, and ${\hat \sigma}_2 = 16 \pi^2 \sigma_2/(d(d+2))$. This set of equations have the following fixed points or lines :
\begin{eqnarray}
({\rm i}) \,\,\, \lambda_2 &=& \lambda_3={\hat \rho}={\hat \rho}_1={\hat \rho}_2={\hat {\overline \rho}}_1={\hat \sigma}={\hat \sigma}_1={\hat \sigma}_2=0, \nonumber \\ 
({\rm ii}) \,\,\, \lambda_3 &+& {\hat \rho}-{\hat \rho}_1 - {\hat \rho}_2 -2 {\hat \sigma}_2 = -8, \,\,\, {\hat \sigma}={\hat \sigma}_1=0, \nonumber \\ 
({\rm iii}) \, \, \, \lambda_3 &+& {\hat \rho}-{\hat \rho}_1 - {\hat \rho}_2 -2 {\hat \sigma}_2 = -5, \,\,\, {\hat \sigma}={\hat \sigma}_1={\hat \sigma}_2=0, \nonumber \\ 
({\rm iv}) \, \, \, \lambda_3 &+& {\hat \rho}-{\hat \rho}_1 - {\hat \rho}_2 -2 {\hat \sigma}_2 = -4, \,\,\, {\hat \sigma}_2=0.
\end{eqnarray}
Among them, the resulting fixed point values of $\lambda_3$ in the cases (ii) and (iii) are found not to satisfy the elastic stability condition, eq.(\ref{stabineq}). In fact, we obtain $\lambda_3=-104/93$ in case (ii) and $-5/3$ in case (iii), respectively. Thus, these cases are unphysical. Further, in the case (iv), we find that ${\hat \rho}$ and ${\hat \sigma}$ are always negative using the elastic stability condition $\lambda_3+1 > 0$. Thus, just as in the similar situation in the isotropic approximation, this case is also judged to be unphysical. In contrast, the linear stability of the nematic fixed point (i) is easily verified. Then, if working around this nematic fixed point, the analysis on the disorder function $R(z)$ is the same as in Ref.\cite{Feldman} and will not be repeated here. Therefore, we reach again the conclusion that the only possible stable fixed point in ${\tilde T} \to 0$ limit is expressed as the nematic one with no vortex-skyrmions. This conclusion that the orbital anisotropy is irrelevant is quite reasonable, judging from the fact that, even in the liquid crystal case \cite{Feldman}, the fixed point expression of the Frank energy (i.e., with $\lambda_2=\lambda_3=0$) is the continuum version of the ferromagnetic Heisenberg spin model with no orbital anisotropy. Further, the above result that, at least at weak disorder, all topological defects can be irrelevant at long distances implies that the superfluid rigidity defined from the current-current correlation function remains finite, because pure Goldstone modes play no roles of destroying superfluidity.

\section{Summary and Discussion}

In this paper, we have shown through calculation of free energy that, in the GL region outside the critical region, the disordered ABM state is lower in free energy than other candidates of an equal-spin pairing state detected as the A-like phase in superfluid $^3$He in aerogel. The local anisotropy characteristic of the aerogel structure plays essential roles in reaching this conclusion, because an anisotropy favors more anisotropic pairing states. If the scattering events are fully isotropic, a much stronger disorder is needed for another ESP state to be realized, although, then, $T_c$ itself would be extremely lowered or vanish. The absence or presence of the genuine long range superfluid order is not essential to a possible change of pairing states: In a situation with a long range order destroyed over some temperature range due to the thermal fluctuation, the entropic term lowers the free energy of some locally ordered state. The vortex liquid regime in the superconducting vortex phase diagram \cite{Nattermann,RI96} is its typical example. Similarly, even in the present case where a static randomness destroys a long range order, a free energy gain from the random-field term overcomes a cost of the elastic (gradient) energy \cite{IM,AI05,Volovik07}. 

In the present work, we have given one possible scenario of the globally isotropic disordered ABM state with a finite superfluid density \cite{Picket}: The A-like phase is an elastic glass \cite{Nattermann} and is in the ABM pairing state with superfluidity as well as in $^3$He in aerogels with an uniaxially stretched anisotropy over large scales [12,9]
. An alternative scenario will be the case in which disorder-induced topological defects including the vortex-skyrmions are pinned by the disorder itself at time scales seen in real experiments. In this case, a nonvanishing superfluid response may be observed. At present, however, it is unclear whether these scenarios assuming globally isotropic samples are relevant to real systems or not. In our opinion, for further development of the present subject, it is necessary for experimentalists to clarify whether globally isotropic aerogel samples are truly available among those used in experiments.

\begin{acknowledgements}
This work is partly supported by a Grant-in-Aid from MEXT of Japan. K.A. is supported by a Grant-in-Aid for JSPS Fellows. 
\end{acknowledgements}

\appendix
\section{} 

In Ref.\cite{AI07}, the strong coupling (SC) corrections, $\delta \beta_j^{({\rm sc})}$ and $\delta {\tilde \beta}_j^{({\rm sc})}$, to the GL-quartic parameters $\beta_j$ were examined in details. Based on the results obtained there, we list here the estimated pressure dependence of $\beta_{\rm N}$ (${\rm N}=$ BW, ABM, P, and R) in Table I. 

\begin{table}
\begin{center}
\begin{tabular}{c c c c c c c c}
\hline
\hline
P[bar]  &  BW & ABM & P(planar) & R(robust) \\
\hline
24 & 1.243 & 1.245 & 1.445 & 1.630 \\
28 & 1.220 & 1.192 & 1.414 & 1.596 \\
34.4 & 1.210 & 1.155 & 1.399 & 1.578 \\
\hline
24 & 1.267 & 1.278 & 1.473 & 1.654 \\
28 & 1.244 & 1.227 & 1.443 & 1.621 \\
34.4 & 1.233 & 1.190 & 1.428 & 1.603 \\
\hline
\hline
\end{tabular}
\caption{$\beta_{\rm N}/\beta_0(T)$ value at $T=T_{c0}$ of each pairing state for $1/(2 \pi T_{c0} \, \tau)=0$ (upper half) and $0.065$ (lower half).}
\end{center}
\end{table}[h]
\\

The data in Table I show that, with increasing disorder, the SC correction in the ABM case is weakened more remarkably compared with those of other pairing states, leading to a rapid narrowing of the temperature range of the A-like phase (see Fig.3). Nevertheless, this effect is not quantitatively substantial at all and does not lead to replacement of the ABM state with other one including the planar or robust state. 

\section{}

In this Appendix, derivation of the free energy of the continuum version of the random Ising spin model 
\begin{eqnarray}
{\overline {\cal S}}_{\rm ising} &=& \int d^3r \sum_{a,b} \biggl[ \delta_{a,b} \biggl( \frac{\tau_0}{2} (\phi^{(a)})^2 + \frac{1}{2}(\nabla \phi^{(a)})^2 
\nonumber \\ 
&+& \frac{g}{4} (\phi^{(a)})^4 \biggr) - \frac{u}{4} (\phi^{(a)} \phi^{(b)})^2 \biggr]
\end{eqnarray}
will be reviewed based on Ref.\cite{Dotsenko}. The analysis proceeds as follows. First, we divide $\phi$ into its mean field, which is $\langle \phi \rangle_{\rm MF}$ in $T < T_c$ and zero in $T > T_c$, and a fluctuation $\delta \phi^{(a)}$. Next, the fluctuation part in ${\overline {\cal S}}_{\rm ising}$ is assumed to be well approximated by the trial action 
\begin{equation}
{\overline {\cal S}}_g = \frac{V}{2} \int_{\bf p} \sum_{a,b} {\cal G}^{-1}_{ab}({\bf p}) \, \delta \phi_a(-{\bf p}) \, \delta \phi_b({\bf p}).
\end{equation}
Then, when calculated according to eq.(\ref{FGVM}), the free energy is well approximated by 
\begin{equation}
\frac{F}{V} = \frac{1}{2n} \int_{\bf p} {\rm tr} \, {\rm ln}({\bf {\cal G}}^{-1}({\bf p})) + \frac{T}{nV} \langle ({\overline {\cal S}}_{\rm ising} - {\overline {\cal S}}_g) \rangle_g
\end{equation}
with taking $n \to 0$ limit at the end, where $V$ is the volume, and ${\rm tr}$ denotes here the trace over the replica indices. Finally, $F$ is calculated in terms of the solution of the saddle-point equations: 
\begin{eqnarray}
\frac{\delta F}{\delta {\cal G}_{aa}({\bf p})} = 0, \\ 
\frac{\delta F}{\delta {\cal G}_{ab}({\bf p})} (a \neq b) = 0, \\
\frac{\delta F}{\delta \langle \phi \rangle_{\rm MF}} = 0.
\end{eqnarray}
The replica-symmetry breaking, which may not be negligible in the critical region \cite{comfinal}, will not be considered for ${\cal G}_{ab}$. Then, we have 
\begin{equation}
{\cal G}_{ab} = {\cal G}_c(p) \delta_{a,b} + \mu \, ({\cal G}_c(p))^2, 
\end{equation}
where 
\begin{equation}
{\cal G}_c(p) = \frac{1}{\lambda + p^2}. 
\end{equation}
The "mass" $\lambda$ of fluctuation $\delta \phi$ will be determined through eq.(B4). The parameter $\mu$ is related to the glass order parameter, which is, by definition \cite{EA}, nonvanishing below $T_c$, and is determined by eq.(B5), while the average value $\langle \phi \rangle_{\rm MF}$ follows from eq.(B6). 
Further, we focus only on the region outside the critical region in which 
\begin{equation}\label{condition}
\biggl(\frac{3g}{2 \pi} \biggr)^2 \ll |\lambda| < 1
\end{equation}
and assume $u \ll g \ll 1$. The latter relation is safely satisfied in superfluid $^3$He at weak static disorder. Then, 
$\mu$ simply becomes $2 u (\langle \phi \rangle_{\rm MF})^2$, and the free energy density $f$ is expressed in the form 
\begin{eqnarray}
f &=&  - \frac{1}{2} [{\rm ln} {\cal G}_c(p) ] + \frac{1}{2}(\tau_0 - \lambda) [ {\cal G}_c(p) ] + \frac{1}{4}(3g -2u) [{\cal G}_c]^2 \nonumber \\
&-& \frac{\lambda}{4} (\langle \phi \rangle_{\rm MF})^2 + \frac{g}{4} (\langle \phi \rangle_{\rm MF})^4, 
\end{eqnarray}
where 
$[F({\cal G}_c(p))] = \int_{\bf p} F({\cal G}_c(p))$. Below, $\lambda$ will be denoted as $\lambda_p$ ($\lambda_f$) in $T > T_c$ ($T < T_c$). The free energy density $f_p$ in $T > T_c$ simply becomes 
\begin{equation}\label{fampp}
f_p = - \frac{\lambda_p^{3/2}}{12 \pi} + \frac{p_c}{4 \pi^2}\lambda_p - \frac{1}{4}(3g -2u) [{\cal G}_c]^2 + f_c
\end{equation}
except a constant $f_c$ depending only on a momentum cutoff $p_c$, where $\lambda_p = \tau_0 + (3g-2u)p_c/(2 \pi^2)$. The first term of $f_p$ is the ordinary Gaussian fluctuation term leading to the singular behavior $\sim (T - T_c)^{-1/2}$ of the specific heat. Under the condition (\ref{condition}), this $\lambda_p^{3/2}$ term may be neglected together with the corresponding one in $f_f$ given below. The free energy density $f_f$ below $T_c$ is $f(\lambda=\lambda_f)$ and takes the form 
\begin{eqnarray}\label{fampf}
f_f &\simeq& - \frac{\lambda_f^{3/2}}{12 \pi} + \frac{p_c}{4 \pi^2} \lambda_f - \frac{3}{4} \lambda_f [{\cal G}_c] - \frac{1}{4}(3g-2u) [{\cal G}_c]^2 
\nonumber \\
&-& \frac{\lambda_f}{4} (\langle \phi \rangle_{\rm MF})^2 
+ \frac{g}{4} (\langle \phi \rangle_{\rm MF})^4 + f_c \nonumber \\
&=& - \frac{\lambda_f^2}{16 g} - \frac{p_c}{8 \pi^2} \lambda_f - \frac{1}{4}(3g-2u) [{\cal G}_c]^2 + f_c, 
\end{eqnarray}
where 
\begin{equation}\label{lamf}
\lambda_f = 2(-\tau_0 - (3g-2u) [{\cal G}_c]), 
\end{equation}
and 
\begin{equation}\label{MFamp}
(\langle \phi \rangle_{\rm MF})^2 = \frac{\lambda_f}{2g}. 
\end{equation}
Although the present analysis takes account of fluctuation effects, the critical region is neglected. Nevertheless, the expressions (\ref{fampp}) and (\ref{fampf}) ensure a continuous transition at $T_c$ defined by $\lambda_p=0= - \lambda_f/2$.

\section{}

In this Appendix, we explain why the planar state is not realized in the GL region. To do this, let us first examine the gradient energy in the planar state. The symmetry variable $a_{\mu,i}$ of the planar pair-field is expressed as 
\begin{equation}
a_{\mu,i} = \frac{1}{\sqrt{2}} R_{\mu,k} \delta^{\rm T}_{k,i} e^{i \Phi},
\end{equation}
where $\delta^{\rm T}_{i,j} = \delta_{i,j} - {\bf l}_i {\bf l}_j$, and $R_{\mu,i}$ is the real rotation matrix expressing the BW state. Below, this ${\bf l}$-vector expressing the local anistropy axis in the planar state will be represented in terms of the same Euler angles as those in the ABM state (see eq.(\ref{lvector})). After substituting eq.(C1) into the gradient energy (\ref{Ssymn}), any term unaccompanied by $\nabla \delta^{\rm T}_{i,k}$ can be neglected in the present harmonic approximation, because the disorder term in ${\overline {\cal S}}_{\rm sym}$ depends only on the Euler angle $\theta_l$ expressing ${\bf l}$. Although a close examination is necessary for a cross term like $R_{\mu,k} (\nabla R_{\mu,m}) \delta^{\rm T}_{l,j} \nabla \delta^{\rm T}_{k,i}$, this term is found to depend only on the Euler angle $\phi_l$ in the present harmonic approximation. Then, the gradient energy related to the disorder term is simply 
\begin{eqnarray}
&+& \frac{|\Delta_{\rm MF}|^2}{4} \int d^3r (K_2 \partial_i \delta^{\rm T}_{k,j} \partial_i \delta^{\rm T}_{k,j} + 2 K_1 \partial_i \delta^{\rm T}_{k,i} \partial_j \delta^{\rm T}_{k,j}) \nonumber \\
&=& |\Delta_{\rm MF}|^2 \frac{K_1+K_2}{2} \int d^3r \biggl( ({\rm div}{\bf l})^2 + (({\bf l}\cdot \nabla) {\bf l})^2 \nonumber \\
&+& \frac{K_2}{K_1+K_2} ({\bf l}\cdot{\rm curl}{\bf l})^2 \biggr).
\end{eqnarray}
By applying the present harmonic approximation to eq.(C2) again, the resulting harmonic elastic energy is found to be 2.4 times bigger than the corresponding one eq.(\ref{Ssymis}) for the ABM case. It means that the free energy gain due to the quenched disorder in the planar state is smaller than eq.(\ref{DFsymA}) in magnitude. Further, since $\beta_{\rm P} > \beta_{\rm A}$, the planar state cannot become stable through $F_{\rm amp}$ (see sec.III). Therefore, no possibility of realizing the planar pairing state due to the impurity disorder is expected anywhere in the phase diagram at least in GL theory.



\begin{thebibliography}{99}
\bibitem{com3} Strictly speaking, a self-average of ${\cal S}_{2,{\rm dis}}$ also appears in ${\overline {\cal S}}$. However, it is already included in $\alpha$ and need not be considered here.
\bibitem{comdipole} Consequently, our neglect, in evaluating free energy, of the dipole energy term suppressing roles of the gradient energy at larger scales is justified.
\end{thebibliography}
\end{document}